\documentclass[prd,preprint,superscriptaddress,preprintnumbers,nofootinbib]{revtex4}
\usepackage{graphicx}
\usepackage{epsfig}
\usepackage{bm}
\usepackage{latexsym,amsmath,amssymb,wasysym,float}
\usepackage{mathrsfs}  
\usepackage{color}
\usepackage{tablefootnote}
\usepackage{enumitem}

\newcommand{\postscript}[2]{\setlength{\epsfxsize}{#2\hsize}
   \centerline{\epsfbox{#1}}}

\usepackage{ dsfont }
\usepackage{float}
\usepackage{enumitem}
\usepackage{mathrsfs}  
\usepackage{hyperref}
\usepackage{multirow}

\usepackage[usenames,dvipsnames]{xcolor}
\definecolor{orange}{cmyk}{0,0.5,1,0}
\definecolor{rossoCP3}{cmyk}{0,.88,.77,.40}
\definecolor{graa}{rgb}{0.8,0.8,0.8}
\definecolor{blaa}{rgb}{0.2,0.2,0.6}

\def\simlt{\mathrel{\lower2.5pt\vbox{\lineskip=0pt\baselineskip=0pt
           \hbox{$<$}\hbox{$\sim$}}}}
\def\simgt{\mathrel{\lower2.5pt\vbox{\lineskip=0pt\baselineskip=0pt
           \hbox{$>$}\hbox{$\sim$}}}}

\newcommand{\be}{\begin{equation}}
	\newcommand{\ee}{\end{equation}}
\newcommand{\ba}{\begin{eqnarray}}
	\newcommand{\ea}{\end{eqnarray}}

 \allowdisplaybreaks

\begin{document}

\title{{\color{rossoCP3} {\bf The Dark Dimension and the Standard Model Landscape}}}

\vspace{-0.2cm}

\author{\bf Luis A. Anchordoqui}

\affiliation{Department of Physics \& Astronomy,\\ Lehman College,  City University
  of New York, NY 10468, USA}



\author{\bf Ignatios Antoniadis}\thanks{LPTHE is the present address.}

\affiliation{School of Natural Sciences, Institute for Advanced Study, Princeton, NJ 08540, USA}

\affiliation{Laboratoire de Physique Th\'eorique et Hautes \'Energies - LPTHE,\\
Sorbonne Universit\'e, CNRS, 4 Place Jussieu, 75005 Paris, France}

\author{\bf Jules Cunat}

\affiliation{Laboratoire de Physique Th\'eorique et Hautes \'Energies - LPTHE,\\
Sorbonne Universit\'e, CNRS, 4 Place Jussieu, 75005 Paris, France}

\begin{abstract}
 \vspace{-0.2cm}
  \noindent We study the landscape of lower-dimensional vacua of the
  Standard Model (SM) coupled to gravity in the presence of  the
  so-called ``dark dimension'' of size $R_\perp$ in the micron range,
  focusing on the validity of the swampland conjecture forbidding the
  presence of non-supersymmetric anti-de Sitter (AdS) vacua in a consistent quantum gravity  theory. We first adopt the working
  assumption that right-handed neutrinos propagate in the bulk, so that
neutrino Yukawa couplings become tiny due to a volume
  suppression, leading to naturally light Dirac neutrinos. We show
  that the neutrino Kaluza-Klein (KK) towers compensate for the
  graviton tower to maintain stable de Sitter (dS) vacua found in the past, but neutrino oscillation
  data set restrictive bounds on $R_\perp$ and therefore the first KK
  neutrino mode is too heavy to alter the shape of the radion
  potential or the required maximum mass for the lightest neutrino to carry dS rather than AdS vacua
 found in the absence of the dark dimension,  $m_{1,{\rm max}}\lesssim 7.63~{\rm meV}$. 
 We also show that a very light gravitino (with 
  mass in the meV range) could help relax the neutrino mass constraint
  $m_{1,{\rm max}} \lesssim 50~{\rm meV}$. The differences for the
  predicted total neutrino mass $\sum m_\nu$ among these two scenarios
  are within reach of next-generation cosmological probes that may 
  measure the total neutrino mass with an uncertainty
  $\sigma (\sum m_\nu) = 0.014~{\rm eV}$. We also demonstrate that the
  KK tower of a very light gravitino can compensate for the graviton
  tower to sustain stable dS vacua and thus right-handed neutrinos can
  (in principle) be locked on the brane. For this scenario,
  Majorana neutrinos could develop dS vacua, which is not possible in
  the SM coupled to gravity.  Finally, we investigate the effects of bulk neutrino masses in
  suppressing oscillations of the 0-modes into the first KK modes to relax the oscillation bound on $R_\perp$.
\end{abstract} 
\maketitle

\section{Introduction}

Far-off in the infrared, well below the electron mass threshold $m_e$, the
structure of the Standard Model (SM) is really simple: it can be characterized by 4 bosonic degrees of freedom (2 from the photon and 2
from the graviton) plus 6 or 12 fermionic degrees of freedom depending
on whether neutrinos are Majorana or Dirac, respectively. The other
mass scale pertinent to the SM infrared world is the cosmological
constant, $\Lambda \sim 10^{-120} M_p^4 \sim  (0.25 \times
10^{-2}~{\rm eV})^4$, where $M_p = 1/\sqrt{8 \pi G}$ is the reduced Planck mass.

Even though we do not know yet the transformation properties of the
neutrinos under particle-antiparticle conjugation (i.e., whether neutrinos are Majorana or Dirac), other sectors of the
worldwide neutrino program have reached precision stage. Data
analyses from short- and long-baseline neutrino oscillation
experiments, together with observations of neutrinos produced by
cosmic rays collisions in the atmosphere and by nuclear fusion
reactions in the Sun, provide the most sensitive insights to determine
the extremely small mass-squared differences. Neutrino oscillation data can
be well-fitted in terms of two nonzero differences $\Delta m^2_{ij}=m^2_i-m^2_j$ between the squares of the masses of
the three ($i=1,2,3$) mass eigenstates $m_i$; namely, $\Delta m^2_{21} = (7.53 \pm 0.18) \times 10^{-5}~{\rm eV}^2$  and $\Delta m^2_{32} = (2.453 \pm 0.033) \times 10^{-3}~{\rm
  eV}^2$ or $\Delta m^2_{32} = (-2.536 \pm 0.034) \times 10^{-3}~{\rm
  eV}^2$~\cite{ParticleDataGroup:2022pth}. In addition, the total
neutrino mass $\sum m_\nu \equiv \sum_{i=1}^3 m_i$ can be determined
(or more restrictively bounded) by analyzing the impact of
cosmological relic neutrinos on the growth of structure
formation. Assuming a $\Lambda$ cold dark matter (CDM) cosmology, {\it Planck} temperature and polarization data lead to
$\sum m_\nu < 0.26~{\rm eV}$, but when the observations of the cosmic microwave
background (CMB) are complemented with those of baryon acoustic
oscillations (BAO)
the bound becomes more restrictive $\sum m_\nu < 0.13~{\rm
  eV}$~\cite{Planck:2018vyg}. Moreover, when CMB + BAO data are supplemented with
supernovae type Ia luminosity distances
 and confronted with determinations of the growth rate parameter the
 upper limit translates to $\sum m_\nu < 0.09~{\rm
   eV}$~\cite{DiValentino:2021hoh}; see also~\cite{Palanque-Delabrouille:2019iyz,diValentino:2022njd}. Putting all of this
together, we arrive at an intriguing experimental fact: the scale of neutrino
masses, $m_i \lesssim 10^{-2}~{\rm eV}$, is not far from that of the observed vacuum energy $\Lambda
\sim m_i^4$. This happenstance could be the carrier of fundamental
information on the possible connections between particle physics, cosmology, and quantum gravity.

The SM coupled to gravity has a unique four-dimensional vacuum (although possible metastable), but it
has long been known that there may also exist lower-dimensional vacua
stabilized by the Casimir energies of particles with masses
$\ll m_e$~\cite{Arkani-Hamed:2007ryu}. Such vacua can have both de Sitter (dS) as well as
anti-de Sitter (AdS) geometries and their nitty-gritty depends
sensitively on the value of neutrino masses. In particular, if all
neutrinos were Majorana and we compactify the low-energy
effective theory down to three or two dimensions, then
AdS SM vacua would appear for any values of neutrino masses consistent
with experiment. It is noteworthy that these lower-dimensional vacua
are virtually indistinguishable from the SM vacuum at distances $\gtrsim
30~\mu{\rm m}$.

A, seemingly different, but in fact closely related subject has been
the development of the Swampland program that lays out a set of constraints to distinguish
effective theories which can be consistently coupled to quantum gravity in the
ultraviolet (UV) from those which cannot~\cite{Vafa:2005ui}. These constraints
have been formulated in the form of swampland conjectures~\cite{Palti:2019pca,vanBeest:2021lhn,Agmon:2022thq}. A well-known swampland conjecture
is the absence of non-supersymmetric (SUSY) AdS vacua supported by
fluxes in a consistent quantum
 gravity theory~\cite{Ooguri:2016pdq}. This conjecture, if correct, implies that if AdS SM lower
 dimensional vacua exist and are stable, then the 4-dimensional SM
 itself could not be completed in the UV. Automatically, the
 conjecture then also implies that the minimal SM setting with
 Majorana neutrinos would be excluded. If neutrinos are Dirac,
 however, the conjecture constrains the mass of the lightest neutrino
 state, $m_i \lesssim \Lambda^{1/4}$~\cite{Ibanez:2017kvh}.\footnote{We
   note in passing that other swampland conjectures applied to the same
   class of lower-dimensional SM vacua lead to similar constraints on neutrino masses~\cite{Gonzalo:2021zsp}.} But of course, to avoid AdS vacua one can always
 extend the mass spectrum of the low-energy effective theory by adding
 fermionic degrees of freedom in the deep infrared
 region, e.g., from a very light gravitino~\cite{Ibanez:2017kvh}. In plain English, Majorana neutrinos, which in the SM are not
 consistent with the bounds from absence of AdS vacua, can be rescued
 by a very light gravitino, keeping the
 attractive see-saw mechanism for neutrino masses active. The requirements to avoid the instability of non-SUSY AdS
 vacua have been established in the so-called light fermion conjecture~\cite{Gonzalo:2021fma}. 
 
 Another interesting aspect of the Swampland program are
 considerations regarding the behaviour of effective theories with a
 cosmological constant. In particular, the distance conjecture~\cite{Ooguri:2006in} when
 generalized to dS space~\cite{Lust:2019zwm} suggests
 that the smallness of dark energy 
 could signal a universe living at the boundary of the
 field space in quantum gravity with a proper distance given by $-\ln|\Lambda|\,,$ in Planck units. A universal feature of these  asymptotic corners in the string landscape of vacua
 is that they predict a light infinite tower of Kaluza-Klein (KK) states whose
 mass $m_{\rm KK}$ is correlated to $\Lambda$. Actually, by combining
 the generalized distance conjecture for dS with observational data,
 the smallness of the cosmological constant and astrophysical
 constraints led to a scenario with one mesoscopic dimension of micron
 scale~\cite{Montero:2022prj}. This extra dimension, dubbed the dark dimension, opens up at
 the scale 
 $m_{\rm KK} \sim \lambda^{-1} \Lambda^{1/4}$ of the tower, where the proportionality 
factor is estimated to be within the range  
 $10^{-4} \lesssim \lambda \lesssim 10^{-1}$. Within this set-up, 
the 5-dimensional Planck scale (or species scale where gravity becomes strong~\cite{Dvali:2007hz,Dvali:2007wp}) 
is $\Lambda_{\rm QG} \sim m_{\rm  KK}^{1/3} \ M_p^{2/3}\simeq 10^9$ GeV.

The dark dimension scenario enjoys a rich phenomenology:
\begin{itemize}[noitemsep,topsep=0pt]
\item  It provides a
natural set up for right-handed neutrinos propagating in the
bulk~\cite{Montero:2022prj}. Within this framework we expect neutrino masses to occur in the range
$10^{-4} < m_i/{\rm eV} < 10^{-1}$, despite the lack of any
fundamental scale higher than $\Lambda_{\rm QG}$. The suppressed neutrino
masses are not the result of a see-saw mechanism, but rather because
the bulk modes have couplings suppressed by the volume of the dark
dimension (akin to the weakness of gravity at long distances)~\cite{Dienes:1998sb,Arkani-Hamed:1998wuz,Dvali:1999cn,Davoudiasl:2002fq,Antoniadis:2002qm}.
\item It encompasses a framework for primordial black
holes~\cite{Anchordoqui:2022txe,Anchordoqui:2022tgp} and KK
gravitons~\cite{Gonzalo:2022jac} to emerge as interesting
dark matter candidates.
\item It also encompasses an interesting framework for studying cosmology~\cite{Anchordoqui:2022svl,vandeHeisteeg:2023uxj} and
astroparticle physics~\cite{Anchordoqui:2022ejw,Noble:2023mfw}.
\item It  provides a profitable arena to accommodate a very
  light gravitino~\cite{Anchordoqui:2023oqm}.
\end{itemize}
In light of this rich phenomenology that connects the various topics
described above, in this paper we examine the landscape of
3-dimensional vacua obtained from compactifying the SM to 3 dimensions
in the presence of the dark dimension. The precise geometry (dS, AdS, or Minkowski) is driven by competing
contributions to the effective lower-dimensional potential. The classical
contributions include the four-dimensional cosmological constant and
the curvature terms resulting from dimensional reduction, while
the quantum contributions are determined by the Casimir energies of SM particles, as
well as of KK excitations of fields propagating in the dark dimension. The fermionic
degrees of freedom we consider in our study are those of left-
and right-handed neutrinos with and without bulk masses, a very light
gravitino, and the KK towers associated to the bulk fields.

The layout of the article is as follows. In Sec.~\ref{sec:2} we review the 3D vacua obtained in the SM coupled
to gravity from the interplay of Casimir forces and the cosmological
constant. In Sec.~\ref{sec:3}
we describe the general structure of the dark dimension scenario, focusing on
the 5D gravity sector along with
the degrees of freedom associated to the 0-modes and their corresponding KK towers that pop up in the 4D
low-energy effective theory. After that, assuming bulk right-handed
neutrinos, in Sec.~\ref{sec:4} we discuss upper limits on the lightest
neutrino mass obtained by balancing bosonic and fermionic degrees of
freedom of the effective radion potential, while imposing at the same time the absence of AdS vacua. In
Sec.~\ref{sec:5} we analyze how the presence of a very light gravitino could
help modify the upper limit on the mass of the lightest neutrino state. In
Sec.~\ref{sec:6} we analyze the effects of bulk neutrino masses in
suppressing oscillations of the 0-modes into the first KK modes, while equalizing the KK bosonic towers to
those associated with the neutrino fields. We reserve Sec.~\ref{sec:7} for our conclusions.

\section{Compactifying the SM on a circle}
\label{sec:2}
 
Consider the action of general relativity (GR)
compactified on a circle of radius $R$, 
\begin{eqnarray}
  S_{\rm GR} & = & \int d^3x d\phi \sqrt{-g_{(4)}} \left( \frac{1}{2}
    M_p^2 {\cal R}_{(4)} - \Lambda_4 \right) \nonumber \\
  & \to & \int d^3x \sqrt{-g_{(3)}} (2 \pi r) \left[\frac{1}{2}
    M_p^2 {\cal R}_{(3)} - \frac{1}{4} \left(\frac{R}{r} \right)^4
          V_{\mu \nu} V^{\mu \nu} - M_p^2 \left(\frac{\partial R}{R}\right)^2 -
    \Lambda_4 \left(\frac{r}{R}\right)^2 \right]\!\!,
\label{SGR}
\end{eqnarray}  
where $g_{(d)}$ is the determinant of the $d$-dimensional metric tensor, ${\cal
  R}_{(d)}$ the $d$-dimensional Ricci scalar, $V_{\mu \nu}$ the field strength of the graviphoton, $0\leq
\phi < 2 \pi$, $\Lambda_4$ is the 4D cosmological constant, and where $r$ is an
arbitrary scale that we fix to the expectation value of the radion
field $R$. For distances larger than $R$, there is an effective 3D theory with metric parametrized by
\begin{equation}
  ds^2_{(4)} = \frac{r^2}{R^2} \ ds^2_{(3)} + R^2 \left( d\phi^2 -
    \sqrt{2}{M_p r} V_\mu dx^\mu \right)\, ,
\label{Wmetric}
\end{equation}
where $V_\mu$ is the graviphoton. From (\ref{SGR}) it
is straightforward to see that the classical potential of the radion coming from the 4D cosmological constant,
\begin{equation}
V_{C}(R) = 2\pi
r \left(\frac{r}{R}\right)^{2} M_p^{2}\Lambda_4 = 2\pi r
\left(\frac{r}{R}\right)^{2} \Lambda \,,
\label{vcee}
\end{equation}
is runaway, and makes the circle decompactify.

Nevertheless, $\Lambda_4$ is so tiny that quantum corrections to
the vacuum energy from
the lightest SM modes could become important to stabilize the radion
potential. The 1-loop corrections to $V_C(R)$ are driven by the Casimir
energy (inferred from loops wrapping the circle) of the lightest SM particles, which are UV insensitive and
have been calculated in~\cite{Arkani-Hamed:2007ryu}. 

Altogether, if we compactify the SM + GR on a circle, the radion gets an effective potential of the form
\begin{equation}
    V(R) =V_C(R)+\sum\limits_{i}V_{i}(R) \,,
\label{V}
\end{equation}
where $V_{i}$ denotes the contribution from the 1-loop Casimir energy of the
particle $i$. For a particle of mass $m_{i}$ with $N_{i}$ degrees of
freedom, the contribution to the potential is given by
\begin{align}
    V_{i}(R)= (-1)^{s_i} \ \frac{N_{i}r^{3}m_{i}^{2}}{4\pi^{3}R^{4}}\sum\limits_{n=1}^{\infty}\frac{K_{2}(2\pi
  Rm_{i}n)}{n^{2}}\cos(n\theta_{i}) \,,
\end{align}
where $s_i = 0 (1)$
for fermions (bosons),  $\theta_i$ is an angle defining the
periodicity around the circle by a phase $e^{2\pi i \theta}$, and
\begin{align}
    K_{\nu}=\frac{1}{2}\int_{0}^{\infty}d\beta\ \beta^{\nu-1}\exp\left[-\frac{z}{2}\left(\beta+\frac{1}{\beta}\right)\right]
\end{align}
is the Bessel function.

For massless
particles, 
\begin{equation}
V_i (R) = 2 \pi r \rho_i (R) \left(\frac{r}{R}\right)^2 \,,
\end{equation}
with 
\begin{align}
    \rho_{i}(R)= (-1)^{s_i} \frac{1}{16\pi^{6}R^{4}}\text{Re}\left[\text{Li}_{4}(e^{i\theta_{i}})\right] \,,
\label{rhoi}
\end{align}
where
\begin{align}
    \text{Li}_{n}(z)=\sum\limits_{k=1}^{\infty}\frac{z^{k}}{k^{n}}
\end{align}
is the polilogarithm. Throughout we consider particles with periodic boundary
conditions; namely $\theta_{i}=0$. A relevant relation is then
$\text{Li}_{n}(1)=\zeta(n)$, where $\zeta(z)$ is the Riemann
zeta-function. Note that for a massive particle $m_i$, the Casimir
energy density is exponentially suppressed by a factor of the form
$\exp(-2\pi Rm_i)$ at large $m_i$, and therefore only the light particles have to be
considered in the sum of (\ref{V}). All in all, in the case of the SM spectrum, we have:
\begin{itemize}[noitemsep,topsep=0pt]
    \item 2 massless bosonic degrees of freedom for the photon,
    \item 2 massless bosonic degrees of freedom for the graviton,
    \item 4 (2) fermionic degrees of freedom for each of the three
      Dirac (Majorana) neutrinos of masses $m_{1}$, $m_{2}$ and
      $m_{3}$.
\end{itemize}
The effective potential (\ref{V}) can be
recast as 
\begin{equation}
V(R) \simeq V_C (R) - 4 \left(\frac{r^3}{720 \pi R^6} \right) +
\sum_{i} \frac{N_i}{720 \pi} \frac{r^3}{R^6} \Theta (R_i - R) \,,
\label{VB}
\end{equation}
where $R_i = 1/m_i$ and $\Theta (x)$ is the step function, with $i=
1,2,3$. Note that in (\ref{VB}) we only take into account (nearly) massless 4D states and look for a 3D vacuum
of toroidal compactification, where the 4D graviphoton is projected
out. Note also that if we only consider the first two terms in the
potential, $V(R)$ develops a maximum at
\begin{equation}
  R_{\rm max} = \left(\frac{1}{120 \pi^2 \Lambda} \right)^{1/4} \simeq
  11~{\rm \mu m}
  \,,
\end{equation}
corresponding to a mass scale
\begin{equation}
  m_{\rm max} = \frac{1}{2 \pi R_{\rm max}} \simeq 2.11~{\rm meV} \,,
\end{equation}  
which is below about the neutrino mass scale. Then, as
the value of $R$ decreases the various neutrino thresholds open up and
sooner or later overwhelm the bosonic contribution to $V(R)$. Thus,
provided $R_i < R_{\rm max}$ the effective radion potential would develop minima.
As can be seen from Eqs.~(\ref{vcee}) and (\ref{VB}) $r$ is an overall normalization
scale which does not influence the nature of AdS or dS vacua, and so
following~\cite{Arkani-Hamed:2007ryu} in our calculations we set $2 \pi r=1~{\rm GeV}^{-1}$.

Before proceeding, we pause to note that 
matter effects provide the only means by which we can determine the
sign of $\Delta m^2_{ij}$. Indeed, because of matter effects in the
Sun, we know that $\Delta m^2_{21}>0$. However, the atmospheric mass splitting $\Delta m^2_{32}$
 is essentially measured only via neutrino oscillations in vacuum and,
 as noted in the Introduction, its sign is unknown. This implies that  as of today it is not possible to decide whether the $\nu_3$
neutrino mass eigenstate is heavier or lighter than the $\nu_1$ and
$\nu_2$ eigenstates. The scenario, in which the $\nu_3$ is heavier, is
referred to as the normal mass hierarchy or normal ordering (NO). The other scenario, in
which the $\nu_3$ is lighter, is referred to as the inverted mass
hierarchy or inverted ordering (IO). 
It has been argued that the latest cosmological constraint, $\sum
m_\nu < 0.09~{\rm eV}$, provides Bayesian evidence for the
NO~\cite{Jimenez:2022dkn}. However, whether Bayesian suspiciousness is
enough to disfavor the IO is still a matter of debate, see
e.g.~\cite{Gariazzo:2022ahe,Gariazzo:2023joe}. Moreover, some
cosmological parameters are correlated with the total neutrino mass,
and so in beyond $\Lambda$CDM models that tend to ameliorate the
Hubble constant tension the bound on $\sum m_\nu$ could be relaxed,
see e.g.~\cite{Sekiguchi:2020igz}. Herein, we will consider the two
possibilities: for NO, we have $m_{1}<m_{2}<m_{3}$, whereas for IO, we have $m_{3}<m_{1}<m_{2}$.

\begin{figure}[!h]
\centering
  \includegraphics[width=15cm]{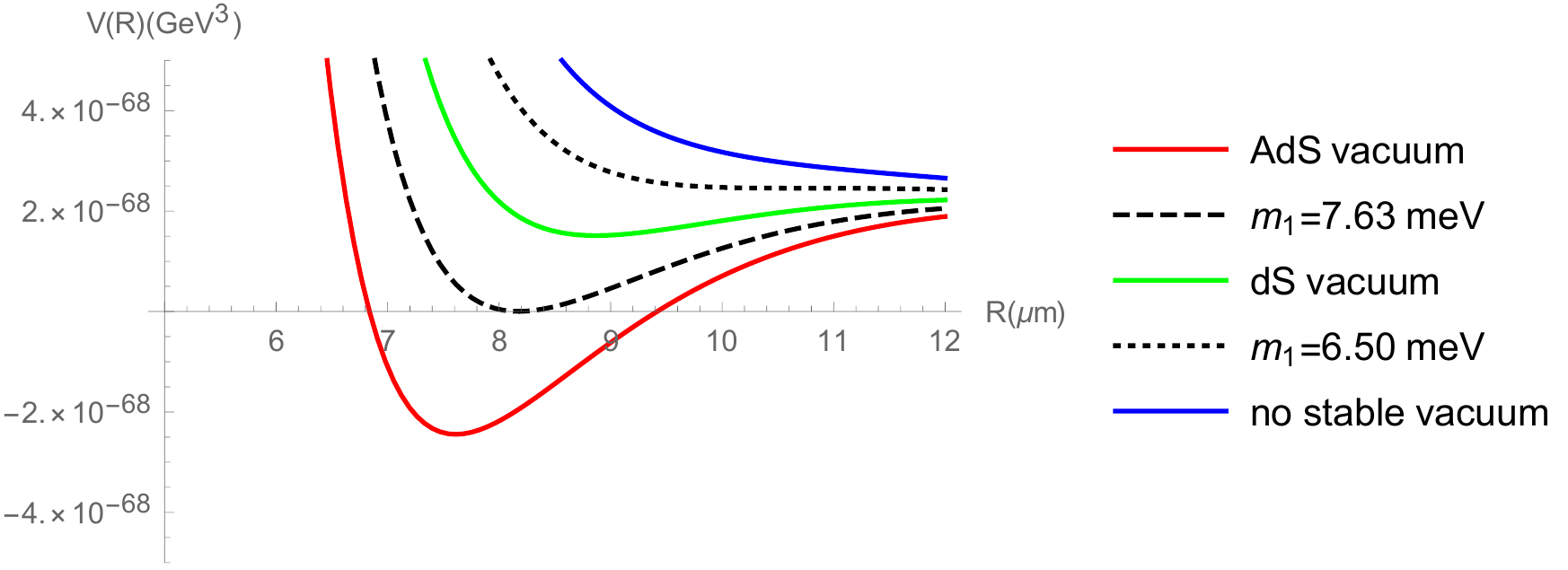}
  \caption{Effective radion potential with Dirac neutrinos and NO.}
  \label{fig:example}
 \end{figure}

Now, depending on the nature and on the masses of the neutrinos we can
obtain different types of SM vacua. As an illustration, in Fig.~\ref{fig:example} we show
the landscape of vacua for Dirac neutrinos with normal ordering. The required maximum mass on the lightest neutrino
state to avoid AdS vacua is $m_{1,{\rm max}} = 7.63~{\rm meV}$. If
neutrinos were Majorana particles, then  AdS vacua would
appear for any values of neutrino masses consistent with
experiment. Hence, the AdS non-SUSY conjecture rejects the case of Majorana
neutrinos if the low energy effective theory is SM + GR.

It is well-known that models in which the observed weakness of gravity at long distances is
due to the existence of compact spatial dimensions~\cite{Arkani-Hamed:1998jmv,Antoniadis:1998ig} provide a
compelling framework to explain Dirac neutrino masses~\cite{Dienes:1998sb,Arkani-Hamed:1998wuz,Dvali:1999cn,Davoudiasl:2002fq,Antoniadis:2002qm}. Hence, it is of
interest to investigate how new degrees of freedom that open up in 5D
models of neutrino physics would modify the
shape of $V(R)$.

\section{Extending the low-energy effective theory}
\label{sec:3}

We consider the compactification framework of~\cite{Arkani-Hamed:1998jmv,Antoniadis:1998ig}, in which it is
natural to assume that, in the case of an orbifold compactification,
the particles charged under the SM gauge group are locked on a
3-brane. Because SM gauge and matter fields live on the brane, the only long-range interaction which sees the extra dimensions is
gravity. Herein, we restrict ourselves to the case of one mesoscopic
extra-dimension characterized by a length-scale in the micron range,
dubbed the dark dimension~\cite{Montero:2022prj}. Actually, null results from
searches of deviations from Newton's gravitational inverse-square law
place an upper bound on the compactification radius, $R_\perp < 30~\mu{\rm m}$~\cite{Lee:2020zjt}.

A point worth noting at this juncture is that a connection has been made elsewhere~\cite{Anchordoqui:2023oqm} between the dark dimension
and the scale of SUSY breaking. As shown in~\cite{Anchordoqui:2023oqm}, the gravitino mass
$m_{3/2}$ and the scale of SUSY breaking can directly be determined
from the dark energy density. Furthermore, as also shown in~\cite{Anchordoqui:2023oqm}, the dark
dimension provides a cost-effective background to host a very light
gravitino. Since this gravitino would modify the mass spectrum of
the effective theory in the deep infrared region, it is of interest to
study how the extra degrees of freedom modify the maximum and minima
of the effective radion potential.

With this in mind, herein we consider 5D supergravity, which contains 8
bosonic degrees of freedom (5 for the graviton and 3 for a gauge
field) and 8 fermionic degrees of freedom ($2 \times 4$ for two
gravitinos). From a 4D
perspective, the degrees of freedom in the bosonic sector are: 2 for
the graviton + 2 for the graviphoton + 1 for the radion + 2 for the gauge field + 1
for an extra scalar. This corresponds to ${\cal N}=2$ SUSY: graviton + vector 
multiplet, each containing one Dirac spinor (2 gravitinos + 2 Weyl
fermions) + their KK excitations. The orbifold breaks SUSY to ${\cal N}=1$. At the massive level the spectrum
is divided by 2 (with cosine and sine wave functions). At the massless
level there is a projection to ${\cal N}=1$ leading to 4 bosonic and 4
fermionic degrees of freedom: spin 2 multiplet + a chiral multiplet
counting the radion, its pseudoscalar partner, and the goldstino. SUSY
breaking makes the gravitino massive by absorbing the goldstino and yields 2 scalars with different masses (of course all are set by
$m_{3/2}$). The content of the gravity multiplet is summarized in Table~\ref{tab:gravitymultiplet}.

In our analysis we first drop the extra scalar assuming it becomes heavier
and keep only the radion. We analyse two scenarios, one in which the
radion is very light and another in which the radion is heavy and does
not partake in carving $V(R)$. Then we consider that the pseudo-scalar
partner of the radion (the axion) is also light. The graviphotons $Z_\mu$ and
$A_\mu^{(0)}$  are taken into account at the massive level, because they have a sine wave
function that vanishes at the brane position and therefore there is no
contribution of the 0-modes to $V(R)$. In Sec.~\ref{sec:4} we will
consider that the gravitino is heavy and plays no role in the shape of
$V(R)$. After that in Sec.~\ref{sec:5} we study the impact of a
very light gravitino on the determination of the effective radion potential.

\begin{table}[!tbh]
   \caption{Gravity multiplet content.}
  \begin{tabular}{c||c|c|c|c|c||c|c|c|c}
    \hline
    \hline
    &\multicolumn{5}{c||}{bosons} & \multicolumn{4}{c}{fermions}\\
    \hline
    \hline
   & \multicolumn{9}{c}{5D~~~~~~~~~~~}\\
   \hline
       ~~~field~~~  & \multicolumn{3}{c|}{$g_{MN}$} &\multicolumn{2}{c||}{$A_{M}$}& \multicolumn{2}{c|}{$\psi_{1,M}$} & \multicolumn{2}{c}{$\psi_{2,M}$} \\
        ~~~spin~~~ &  \multicolumn{3}{c|}{2} &\multicolumn{2}{c||}{1}& \multicolumn{2}{c|}{3/2} & \multicolumn{2}{c}{3/2} \\
        dof &  \multicolumn{3}{c|}{5} &\multicolumn{2}{c||}{3}& \multicolumn{2}{c|}{4} & \multicolumn{2}{c}{4} \\
       $m$ &  \multicolumn{3}{c|}{0} &\multicolumn{2}{c||}{0}& \multicolumn{2}{c|}{0} & \multicolumn{2}{c}{0} \\
\hline
    \multicolumn{10}{c}{$\downarrow$ compactification on $\mathcal{S}^{1}$ $\downarrow$}\\
        \hline
        & \multicolumn{9}{c}{4D 0-modes ($\mathcal{N}=2$)~~~~~~~~~~~}\\
        \hline
        field & $g_{\mu\nu}^{(0)}$ & ~~$Z_{\mu}$~~ & ~~$R_{\perp}$~~ &
                                                                       ~~$A_{\mu}^{(0)}$~~ & ~~$A_{\perp}$~~ & ~~$\psi_{1,\mu}^{(0)}$~~ & ~~$\psi_{1,\perp}$~~ & ~~$\psi_{2,\mu}^{(0)}$~~ & ~~$\psi_{2,\perp}$~~ \\
        spin & 2 & 1 & 0 & 1 & 0 & 3/2 & 1/2 & 3/2 & 1/2 \\
        dof & 2 & 2 & 1 & 2 & 1 & 2 & 2 & 2 & 2 \\
        $m$ & 0 & 0 & 0 & 0 & 0 & 0 & 0 & 0 & 0 \\
        \hline
        &\multicolumn{9}{c}{4D KK-modes $(n\in\mathds{N}^{*})$~~~~~~~~~~~}\\
        \hline
        field & $g_{\mu\nu}^{(n)}$ &  &  & $A_{\mu}^{(n)}$ &  &$\psi_{1,\mu}^{(n)}$ &  & $\psi_{2,\mu}^{(n)}$ &  \\
        spin & 2 &  &  & 1 &  & 3/2 &  & 3/2 &  \\
        dof & 5 &  &  & 3 &  & 4 &  & 4 &  \\
        $m$ & ~~$n/R_\perp$~~ &  &  & ~~$n/R_\perp$~~ &  & ~~$n/R_\perp$~~ &  & ~~$n/R_\perp$~~ &  \\
        \hline
   \multicolumn{10}{c}{$\downarrow$ with the action of $\mathds{Z}_{2}$ and SUSY breaking $\downarrow$}\\
        \hline
        & \multicolumn{9}{c}{4D 0-modes ($\mathcal{N}=1$)~~~~~~~~~~~}\\
        \hline
        field & $g_{\mu\nu}^{(0)}$ &  & $R_{\perp}$ &  & $A_{\perp}$ &$\psi_{1,\mu}^{(0)}$ &  &  &  \\
        spin & 2 &  & 0 &  & 0 & 3/2 &  &  & \\
        dof & 2 &  & 1 &  & 1 & 4 &  &  &  \\
        $m$ & 0 &  & $m_{R_{\perp}}$ &  & $m_{A_{\perp}}$ & $m_{3/2}$ &  &  &  \\
        \hline
        &\multicolumn{9}{c}{4D KK-modes $(n\in\mathds{N}^{*})$~~~~~~~~~~~}\\
        \hline
        field & $g_{\mu\nu}^{(n)}$ &  &  & $A_{\mu}^{(n)}$ &  &$\psi_{1,\mu}^{(n)}$ &  & $\psi_{2,\mu}^{(n)}$ &  \\
        spin & 2 &  &  & 1 &  & 3/2 &  & 3/2 &  \\
        dof & 5 &  &  & 3 &  & 4 &  & 4 &  \\
        $m$ & $n/R$ &  &  & $n/R$ &  & ~~$n/R+m_{3/2}$~~ &  & ~~$n/R+m_{3/2}$~~ &  \\
        \hline
        \hline
    \end{tabular}    
    \label{tab:gravitymultiplet}
\end{table}

\section{Bulk right-handed neutrinos}
\label{sec:4}

Throughout this section we proceed on the working assumptions that
gravitinos (and the SUSY mass spectrum) are heavy and that
neutrino masses derive from 
three 5D fermion fields $\Psi_\alpha \equiv
(\psi_{\alpha L},\psi_{\alpha R})$, which 
are SM singlets and interact on our brane with the three active left-handed neutrinos
  $\nu_{\alpha L}$ and the Higgs doublet preserving lepton number, where the indices $\alpha = e, \mu,\tau$ indicate the
  generation~\cite{Dienes:1998sb,Arkani-Hamed:1998wuz,Dvali:1999cn,Davoudiasl:2002fq,Antoniadis:2002qm}. From
  the viewpoint of 4D observers on the brane, each of the singlet
  fermion fields can be decomposed into
  an infinite tower of KK states, $\psi^\kappa_{L(R)}$,
  with $\kappa = 0,\pm 1, \cdots, \pm \infty$. The right-handed fields $\psi_R^\kappa$ combine with the left-handed bulk states
  $\psi_L^\kappa$ to assemble Dirac
mass terms, which come from the quantized internal momenta in the
dark dimension. In addition, there is a mixing between the bulk states and the active
left-handed neutrinos through Dirac-like mass terms. Note that the
bulk fields can be redefined as $\nu^{(0)}_{\alpha R} \equiv\psi^{(0)}_{\alpha R}$ and
$\nu^{(n)}_{\alpha L(R)}\equiv \Big(\psi^{(n)}_{\alpha
  L(R)}+\psi^{(-n)}_{\alpha L(R)}\Big)/\sqrt 2$, and so after electroweak
symmetry breaking the
mass terms of the Lagrangian read
\begin{eqnarray}
\mathscr{L_{\text{mass}}}  &= & \displaystyle
\sum_{\alpha,\beta}m_{\alpha\beta}^{D}\left[\overline{\nu}_{\alpha
    L}^{\left(0\right)}\,\nu_{\beta R}^{\left(0\right)}+\sqrt{2}\,
  \sum_{n=1}^{\infty}\overline{\nu}_{\alpha
    L}^{\left(0\right)}\,\nu_{\beta
    R}^{\left(n\right)}\right]  
 +  \sum_{\alpha}\sum_{n=1}^{\infty}\displaystyle
m_n \, \overline{\nu}_{\alpha L}^{\left(n\right)} \,
\nu_{\alpha R}^{\left(n\right)} 
                                \, + {\rm h.c.} \nonumber \\
& = & \sum_{i=1}^3 \bar{\mathbb{N}}_{iR} \  \mathbb
      M_i \ \mathbb{N}_{iL}+ {\rm h.c.}  \,,
      \label{calL}
\end{eqnarray}
where $m_{\alpha \beta}^{D}$ is a Dirac mass
matrix, $m_n = n/R_\perp = n m_{\rm KK}$,
\begin{eqnarray}
{\mathbb{N}_{i L(R)}}=\Big(\nu_i^{(0)},\nu_i^{(1)},\nu_i^{(2)},\cdots\Big)^T_{L(R)},
  ~~~~~{\rm{and}}~~~~~ \mathbb M_i=
\begin{pmatrix}
m_i^D&0&0&0&\ldots\\
\sqrt{2}m_i^D&1/R_\perp&0&0&\ldots\\
\sqrt{2}m_i^D&0&2/R_\perp&0&\ldots\\
\vdots&\vdots&\vdots&\vdots&\ddots
\end{pmatrix},
\end{eqnarray}
and where $m_i^D$ are the elements of the diagonalized Dirac mass
matrix $=\mathrm{diag}(m^D_1,m^D_2,m^D_3)$. Greek indices from the
beginning of the alphabet run over the 3 active flavors ($\alpha,\beta
= e,\mu,\tau$), Roman lower case indices over the 3 SM families $(i =
1,2,3)$, and $n$ over the KK modes ($n =1,2,3,..., +\infty$). Note that
$\psi_{\alpha L}^{(0)}$ is projected out from the orbifold. For the
configuration at hand,
\begin{equation}
m_i^D = \frac{y_i v}{\sqrt{\pi R_\perp \Lambda_{\rm QG}}}  \,,
\label{in3}
\end{equation}
where $y_i$'s are the 5-dimensional Yukawa couplings localized on the
SM brane and where $v = 246/\sqrt{2}~{\rm GeV} = 174~{\rm GeV}$ is the Higgs
vacuum expectation value. For the neutrinos, we therefore have 4
fermionic degrees of freedom for each $n\in\mathds{N}$ and for each
$i=1,2,3$ leading to three towers of neutrinos of masses $m_{i}^{(n)}=\lambda_{i}^{(n)}/R_{\perp}$ where $\lambda_{i}^{(n)}$ are solutions of the transcendental equation~\cite{Dienes:1998sb,Arkani-Hamed:1998wuz,Dvali:1999cn,Davoudiasl:2002fq,Antoniadis:2002qm}
\begin{align}\label{transcendentalnewnew}
    \lambda_{i}^{(n)}-\pi\big(m_{i}^{\text{D}}R_{\perp}\big)^{2}\text{cot}\left(\pi\lambda_{i}^{(n)}\right)=0
  \, .
\end{align}

Next, we make contact with experiment to develop some sense for the
orders of magnitude involved. Bearing that in mind, we impose the
\emph{cosmological constraint} on the sum of neutrinos masses. More
precisely, on the sum of the three 0-modes of the three neutrino
towers, 
\begin{align}
    \frac{\lambda_{1}^{(0)}}{R_{\perp}}+\frac{\lambda_{2}^{(0)}}{R_{\perp}}+\frac{\lambda_{3}^{(0)}}{R_{\perp}}<
  \sum m_\nu \, .
\label{cosmoC}
\end{align}
In addition, we
consider the $\Delta m_{ij}^2$ measured by neutrino oscillation
experiments. We remind the reader that $\Delta m_{ij}^2$  are the mass
squared differences between the three 0-modes of the three neutrinos
towers. For NO,  we can write
\begin{equation}
\left(\lambda_{2}^{(0)}\right)^{2} =R_{\perp}^{2}\Delta m_{21}^{2}+\left(\lambda_{1}^{(0)}\right)^{2}
\end{equation}
and
\begin{equation}
\left(\lambda_{3}^{(0)}\right)^{2} =R_{\perp}^{2}\Delta m_{32}^{2}+R_{\perp}^{2}\Delta m_{21}^{2}+\left(\lambda_{1}^{(0)}\right)^{2}.
\end{equation}
Note that the $\Delta m_{ij}^2$'s constrain the parameter space due to
the fact that we need the $\lambda_i^{(0)}$'s to be smaller than $1/2$
in order to have solutions of \eqref{transcendentalnewnew}. By
imposing the $\Delta m_{ij}^2$ constraint we arrive at
\begin{equation}
 \left(\lambda_{1}^{(0)}\right)^{2} <  \frac{1}{4} \label{nu1},
               \end{equation}
               \begin{equation}
R_{\perp}^{2}\Delta
m_{21}^{2}+\left(\lambda_{1}^{(0)}\right)^{2} < \frac{1}{4} \label{nu2},
\end{equation}
and
\begin{equation}
    R_{\perp}^{2}\Delta m_{32}^{2}+R_{\perp}^{2}\Delta
    m_{21}^{2}+\left(\lambda_{1}^{(0)}\right)^{2} < \frac{1}{4} \label{nu3}.
\end{equation}
Combining (\ref{cosmoC}) with \eqref{nu1}, \eqref{nu2}, and \eqref{nu3} we obtain
\begin{align}\label{cosmoplusosc}
\lambda_{1}^{(0)}+\sqrt{R_{\perp}^{2}\Delta
  m_{21}^{2}+\left(\lambda_{1}^{(0)}\right)^{2}}+\sqrt{R_{\perp}^{2}\Delta
  m_{32}^{2}+R_{\perp}^{2}\Delta
  m_{21}^{2}+\left(\lambda_{1}^{(0)}\right)^{2}}< R_{\perp} \sum m_\nu.
\end{align}

\begin{figure}[htb!]
    \postscript{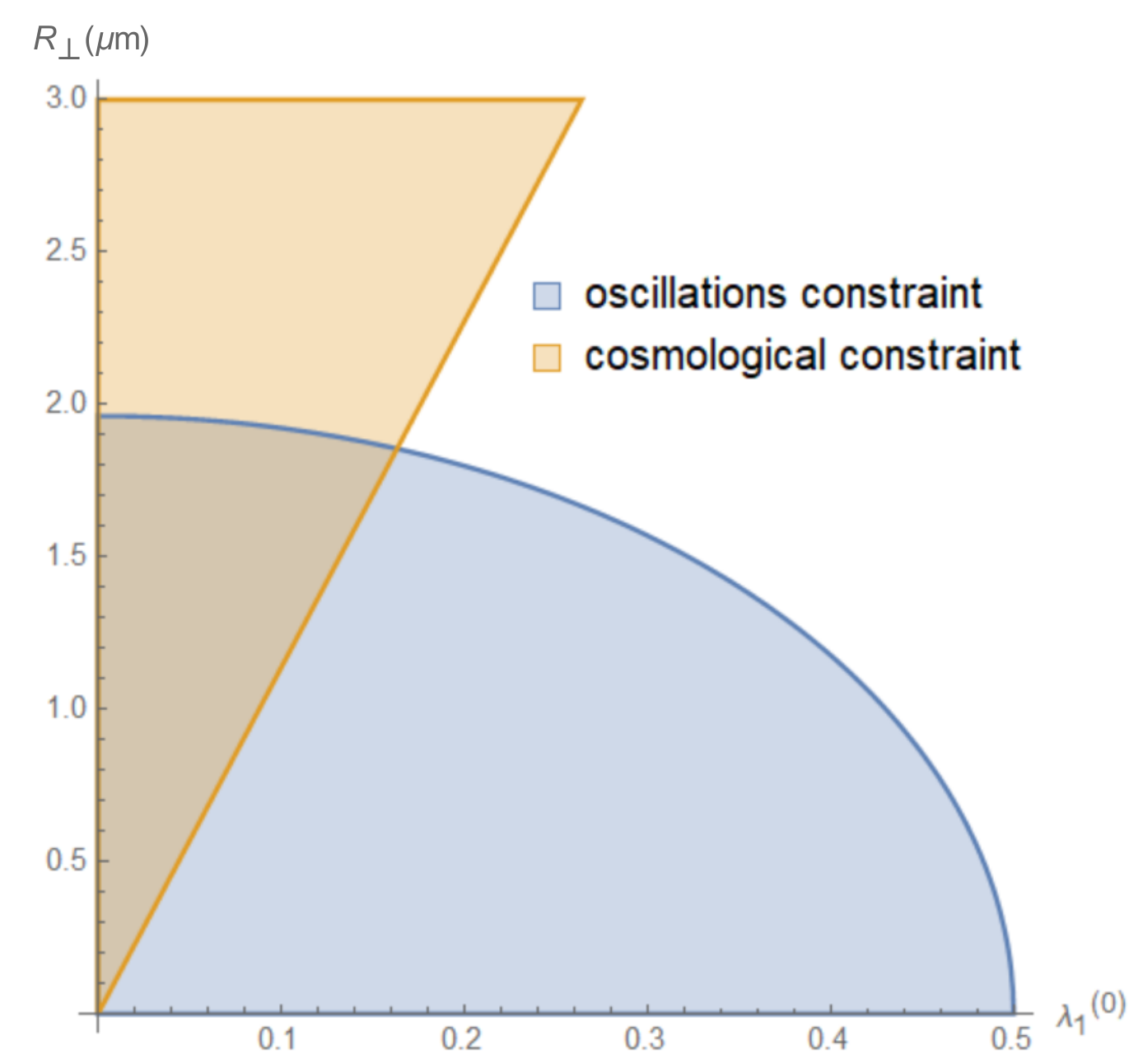}{0.4}
  \caption{Constraints imposed by measurements of $\Delta m^2_{ij}$ in
    oscillation experiments and the inferred $\sum m_\nu$ for NO via
    cosmological observations.}  \label{regions}
  \end{figure}

For the IO case, it is more convenient to keep $\lambda_{3}^{(0)}$ instead of $\lambda_{1}^{(0)}$ (because $\nu_{3}$ is then the lightest neutrino) and the oscillation constraint is now
\begin{equation}
  \left(\lambda_{3}^{(0)}\right)^{2} < \frac{1}{4} \label{nu3inv},
\end{equation}
\begin{equation}
 -R_{\perp}^{2}\Delta m_{21}^{2}-R_{\perp}^{2}\Delta
m_{32}^{2}+\left(\lambda_{3}^{(0)}\right)^{2} <
\frac{1}{4} \label{nu1inv},
\end{equation}
and
\begin{equation}
   -R_{\perp}^{2}\Delta m_{21}^{2}+\left(\lambda_{3}^{(0)}\right)^{2} < \frac{1}{4} \label{nu2inv}.
 \end{equation}
Combining (\ref{cosmoC}) with  \eqref{nu2inv}, \eqref{nu1inv}, and
\eqref{nu3inv} we obtain
\begin{align}\label{cosmoplusoscinv}
\lambda_{3}^{(0)}+\sqrt{-R_{\perp}^{2}\Delta
  m_{32}^{2}+\left(\lambda_{3}^{(0)}\right)^{2}}+\sqrt{-R_{\perp}^{2}\Delta
  m_{21}^{2}-R_{\perp}^{2}\Delta
  m_{32}^{2}+\left(\lambda_{3}^{(0)}\right)^{2}}< R_{\perp} \sum m_\nu
  \, .
\end{align}
Constraints on the $\lambda_{1}^{(0)}-R_{\perp}$ plane for the NO are
encapsulated in Fig.~\ref{regions}. To estimate the allowed region of the parameter
space we have adopted the $\Lambda$CDM cosmological constraint, $\sum
m_\nu < 0.09~{\rm eV}$. As previously noted, the $\Lambda$CDM
cosmological constraint is in tension with the IO.

A more restrictive constraint arises from oscillations of the active
0-modes into the first KK states of the towers. Such a disappearance effect has not
been observed by neutrino oscillation experiments and so experimental data place 
a bound on the compactification radius, $R_\perp < 0.1~\mu{\rm
  m}$~\cite{Machado:2011jt,Forero:2022skg}. This in turn 
implies that the first KK modes in the towers have a mass of at least
${\cal O} (10~{\rm eV})$. Consequently,  the KK excitations are too massive to
counterbalance the effect of the massless bosons, only affecting the radion
potential for very small $R$. As a result, within this set up we
would expect that the
contraints on the maximum mass of the lightest neutrino coincide with
those predicted by the SM + GR. Actually, the constraint on the neutrino mass should
become stronger in the
presence of a light radion field.

We have scanned the parameter space
varying the radion mass $m_{R_\perp}$ and  $m_1$ assuming the NO of 
neutrino masses and assuming the axion is heavy. If we further assume that the radion is heavy and does not
partake in carving $V(R)$, then for $m_1 > 7.63~{\rm meV}$ an AdS
vacuum is formed, whereas for $6.50 < m_1/{\rm meV} < 7.63$ a dS
vacuum is obtained, and if $m_1 < 6.50~{\rm meV}$ there is no
vacuum. If we instead consider the opposite limit in which the radion
is almost massless we find that if $m_1 > 5.28~{\rm meV}$ an AdS
vacuum is formed, while for $4.36 < m_1/{\rm meV} < 5.28$ a dS
vacuum is obtained, and if $m_1 < 4.36~{\rm meV}$ there is no
vacuum. At this point a reality check is in order. Substituting the
maximum mass of the lightest neutrino that can avoid an AdS vacuum
($m_{1, max} = 7.63~{\rm meV}$) and the upper limit of the compactification
radius ($R_\perp = 10~{\rm
  eV}$) into (\ref{in3}) and (\ref{transcendentalnewnew}) we obtain
$y_1 \sim 10^{-4}$. 

We have also scanned the parameter space
varying $m_{R_\perp}$ and  $m_3$, but considering the 
IO of neutrino masses. In this case, if the radion is heavy the values of $m_3$ demarcating
the transitions between geometries with no vacuum, with
dS vacua, and with AdS
vacua are as follows: {\it (i)}~for $m_3 > 2.51~{\rm meV}$ an AdS
vacuum is formed; {\it (ii)}~for $2.04 < m_3/{\rm meV} < 2.51$ a dS
vacuum is obtained; {\it (iii)} for $m_1 < 2.04~{\rm meV}$ there is no
vacuum. On the other hand, if the radion is light and becomes relevant in the
determination of the critical points of $V(R)$, then the lightest
neutrino must be massless, and the minimum radion
mass to avoid an AdS vacuum is $m_{R_\perp} = 25.09~{\rm meV}$,
whereas to support a dS vacuum $m_{R_\perp} < 27.88~{\rm meV}$.

Next, in line with our stated plan,
  we consider the case of NO Dirac neutrinos, assuming a massless
  radion and massless axion. For $m_1 > 2.82~{\rm meV}$ an AdS vacuum
  is formed, while for $1.90 < m_1/{\rm meV} < 2.82$ a dS vacuum is
  obtained, and for $m_1 < 1.90~{\rm meV}$  there is no stable vacuum.
 
In summary, if we assume that righ-handed neutrinos propagate in the
bulk (so that the Yukawa couplings become tiny because of a volume
  suppression) then their KK towers can compensate for the
  graviton tower to avoid AdS vacua. However, neutrino oscillation
  data set restrictive bounds on $R_\perp$ and therefore the first KK
  neutrino mode is too heavy to alter the shape of the radion
  potential or $m_{1,{\rm max}}$ from those predicted by the SM + GR when
  compactified down to 3D.

In closing we note that in Table~\ref{tab:gravitymultiplet} and in our general
  presentation of the mass spectrum in Sec.~\ref{sec:3}, we made the
assumption that the modulino is the goldstino; i.e., it is part of the
massive zero mode of the gravitino. At this stage, it is worthwhile to
point out that the above
consideration is actually model dependent
as e.g., the goldstino could be the fermion of a chiral multiplet if
we have F-term SUSY breaking. In models with high-scale SUSY breaking
the gravitino and the modulino are heavy~\cite{Anchordoqui:2023qxv}. However, in
  models with low-scale SUSY breaking, the modulino could be very
  light. An interesting scenario emerges if the modulino is almost
  massless. On the one hand, if the radion and the axion are also
  almost massless, the modulino fermionic degrees of freedom get
  cancelled by the bosonic degrees of freedom of the radion and axion,
  and the shape of the potential is the same as that considering a
 heavy goldstino with the
  radion and axion also as heavy particles.  On the other hand, it coud be that
  the radion and the axion are heavy. If this were the case, for
  NO of Dirac neutrinos the existence of AdS vacua would be avoided
  if $m_1 < 14.49~{\rm meV}$ and for IO if $m_3 < 11.14~{\rm
    meV}$. This scenario (with a massless modulino and heavy radion and
  axion) also allows to avoid AdS vacua if neutrinos are Majorana particles: for NO the presence of AdS vacua can be avoided if $m_1
  < 9.61~{\rm meV}$ and for IO if $m_3 < 3.43~{\rm meV}$.
  
\section{A very light gravitino}
\label{sec:5}

In line with our stated plan, we now turn to consider the addition of
a very light
gravitino in the mass spectrum. Before proceeding, we pause to note
that various mechanisms have been suggested for SUSY breaking, which span a wide range of gravitino masses: very light,
light, and heavy; a review can be found
e.g. in~\cite{Haber:1984rc}. For gauge-mediated SUSY breaking~\cite{Giudice:1998bp}, with scale $M_{\rm SUSY}
=10~{\rm TeV}$, the minimum gravitino mass is $m_{3/2} \sim 0.1~{\rm
  eV}$~\cite{Anchordoqui:2023oqm}. However, scenarios with tiny masses have also been considered
in the literature, see e.g.~\cite{Qiu:2022lyn}. Herein we adopt the lower bound
on the gravitino mass coming from the LHC  experiment,  $m_{3/2} \gtrsim 1~{\rm meV}$~\cite{ATLAS:2015qlt}. Whichever point of view one may find more convincing, it seems most
conservative at this point to depend on experiment (if possible) to
resolve the issue. 

\subsection{Cosmological inference of the long-distance effective field theory}
\label{sec:5a}

If the gravitino is very light then it would contribute with fermionic
degrees of freedom to the sum in (\ref{VB}) and can help relaxing the bound on
$m_{1,max}$. The different mass scales of $m_1$ (or $m_3$), $m_{3/2}$, and
$m_{R_\perp}$ are summarized in
Table~\ref{tab:radiongravitino}. It is of interest to see whether the
modifications induced on $V(R)$ by the gravitino contribution can be
discerned by future cosmological probes measuring $\sum m_\nu$.

\begin{table}[!tbh]
   \caption{Maximum gravitino mass necessary to avoid an
      AdS vacuum for Dirac neutrinos.\protect\footnote{Impossible means that
      there is always a stable AdS vacuum and whatever 
      that there is no constraint.}}
    \begin{tabular}{p{0.3cm}c|c|c||c|c|c}
\hline
      \hline
      &
        \multicolumn{3}{c||}{NO} & \multicolumn{3}{c}{IO}\\
    \cline{1-7}
     &   $m_{1}$ & $m_{R_\perp}$ & $m_{3/2}$ &  $m_{3}$ & $m_{R_\perp}$ & $m_{3/2}$ \\
        \hline
        \hline
                          &  ~~~~50 meV~~~~ & ~~~~no radion~~~~ & ~~~~2.51 meV~~~~ & ~~~~30 meV~~~~ & ~~~~no radion~~~~ & ~~~~3.42 meV~~~~\\
      &  40 meV & no radion & 2.98 meV & 25 meV & no radion & 3.94 meV\\
     &   30 meV & no radion & 3.77 meV &20 meV & no radion & 4.72 meV\\
     &   20 meV & no radion & 5.39 meV & 15 meV & no radion & 6.02 meV\\
       & 15 meV & no radion & 7.19 meV &  10 meV & no radion & 8.77 meV\\
      &  10 meV & no radion & 12.16 meV &5 meV & no radion & 18.65 meV\\
      &  5 meV & no radion & whatever  &  1 meV & no radion & whatever\\
      & 0 meV &no radion & whatever & 0 meV & no radion  & whatever \\
      \cline{2-7}
      &  50 meV & massless & impossible & 30 meV & massless & impossible\\
      &  40 meV & massless & impossible & 25 meV & massless & impossible \\
      &  30 meV & massless & impossible & 20 meV & massless &impossible\\
      &  20 meV & massless & 1.86 meV & 15 meV & massless & 2.67 meV \\
      &  15 meV & massless & 4.12 meV & 10 meV & massless & 5.38 meV \\
      &  10 meV & massless & 7.56 meV & 5 meV & massless & 10.59 meV \\
      &  5 meV & massless & whatever & 1 meV & massless & 17.94 meV\\
      & 0 meV & massless & whatever & 0 meV & massless & 18.60 meV\\
        \hline
        \hline
       
    \end{tabular}   
    \label{tab:radiongravitino}
\end{table}

Future observations from the Simons
Observatory~\cite{SimonsObservatory:2019qwx}, when complemented with
BAO from DESI~\cite{DESI:2016fyo}
and Rubin LSST weak lensing data~\cite{LSSTDarkEnergyScience:2018jkl}, will allow a determination of the
total neutrino mass with an uncertainty $\sigma (\sum m_i) = 40~{\rm
  meV}$, and with expected improvements in
the determination of the optical depth the sensitivity will refine to
$\sigma (\sum m_i) = 20~{\rm meV}$~\cite{Adhikari:2022sve}. Future measurements of the lensing power spectrum (or cluster
abundances) by CMB-S4~\cite{Abazajian:2019eic}, when supplemented with
BAO from DESI and the {\it Planck}  
measurement of the optical depth, will
provide a constraint on the sum of neutrino masses at the level
$\sigma (\sum m_\nu) = 24~{\rm meV}$, and with expected improvements in
the determination of the optical depth the sensitivity will refine to
$\sigma (\sum m_\nu) = 14~{\rm meV}$~\cite{Chang:2022tzj}. The proposed Probe
of Inflation and Cosmic Origins (PICO), in combination with BAO from
DESI (or Euclid) will reach a sensitivity of $\sigma (\sum m_\nu) =
14~{\rm meV}$~\cite{NASAPICO:2019thw}. This implies that CMB-S4 and the proposed
CMB satellite PICO will be sensitive to a $4\sigma$ detection of the minimum sum predicted by the NO. Far into the
future, measurements of the gravitational lensing of the CMB and the
thermal and kinetic Sunyaev-Zel'dovich effect on small scales by the millimetre-wave
survey CMB-HD may reach a
sensitivity of $\sigma (\sum m_\nu) = 13~{\rm meV}$, corresponding to a $5\sigma$
detection on the sum of the neutrino masses~\cite{CMB-HD:2022bsz}. 

Altogether, this suggests that future cosmological observations will be
able to pin down whether the mass of the lightest neutrino is $m_1 >
7.63~{\rm meV}$ and at the same time will inform us about the possible
existence of a
very light gravitino or other fermionic degrees of freedom in the
deep infrared region.

\subsection{Neutrinos locked on the brane}

If the gravitino is very light, neutrinos could, in principle, be locked on the brane. If
this were the case, it is of interest to
 investigate whether the gravitino could help Majorana neutrinos to
 avoid the existence of AdS vacua. As a first step of this
 investigation we assume neutrinos are Majorana and duplicate the scanning procedure carried out for
 Dirac neutrinos to establish the mass scales of $m_1$ (or $m_3$),
 $m_{3/2}$, and $m_{R_\perp}$ that can avoid AdS vacua. The results are summarized in Table~\ref{tab:radiongravitinomaj}.

\begin{table}[!h]
\caption{Maximum gravitino mass necessary to avoid an
      AdS vacuum for Majorana neutrinos.\protect\footnote{Impossible means that
      there is always a stable AdS vacuum.}}
    \begin{tabular}{p{0.3cm}c|c|c||c|c|c}
      \hline
      \hline
      &\multicolumn{3}{c||}{NO} & \multicolumn{3}{c}{IO}\\
    \cline{1-7}
     &   $m_{1}$ & $m_{R_\perp}$ & $m_{3/2}$ &  $m_{3}$ & $m_{R_\perp}$ & $m_{3/2}$ \\
        \hline
        \hline
                          &  ~~~~50 meV~~~~ & ~~~~no radion~~~~ & ~~~~2.31 meV~~~~ & ~~~~30 meV~~~~ & ~~~~no radion~~~~ & ~~~~3.05 meV~~~~\\
      &  40 meV & no radion & 2.72 meV & 25 meV & no radion & 3.45 meV \\
     &   30 meV & no radion & 3.36 meV &20 meV & no radion & 4.02 meV\\
     &   20 meV & no radion & 4.56 meV & 15 meV & no radion & 4.87 meV\\
       & 15 meV & no radion & 5.67 meV  &  10 meV & no radion & 6.28 meV \\
      &  10 meV & no radion & 7.71 meV &5 meV & no radion & 8.76 meV \\
      &  5 meV & no radion & 12.21 meV  &  1 meV & no radion & 10.97 meV \\
      & 0 meV &no radion & 18.86 meV  & 0 meV & no radion  &  11.14 meV\\
      \cline{2-7}
      &  50 meV & massless & impossible & 30 meV & massless & impossible\\
      &  40 meV & massless & impossible  & 25 meV & massless & impossible \\
      &  30 meV & massless & impossible & 20 meV & massless & impossible\\
      &  20 meV & massless & impossible & 15 meV & massless & impossible\\
      &  15 meV & massless & 2.13 meV  & 10 meV & massless & 2.69 meV\\
      &  10 meV & massless & 4.49 meV & 5 meV & massless &  5.17 meV\\
      &  5 meV & massless & 7.67 meV & 1 meV & massless & 6.81 meV\\
      & 0 meV & massless & 10.84 meV & 0 meV & massless & 6.93 meV\\
        \hline
        \hline
    \end{tabular}
    \label{tab:radiongravitinomaj}
\end{table}

Now, within this scenario neutrinos do not have KK towers, but to
avoid AdS vacua we still have to compensate for the bosonic towers
from the gravity multiplet (whose components propagate into the
bulk). As noted in Sec.~\ref{sec:3} we have adopted $\mathcal{N}=2$
(broken) SUSY in the bulk, namely 8 degrees of freedom for each layer
of the gravity towers; see Table \ref{tab:gravitymultiplet}. Note that
the gravitino tower is shifted from the bosonic towers by
$m_{3/2}$. As a consequence, if this shift is too big, the bosonic
modes could create stable AdS vacua. Now, because of the orbifold
compactification, some of the 0-modes can be projected out. The most
natural choice is to consider only the graviton and the radion 0-modes
among the bosons, and of course the gravitino. We also assume that the
scalar superpartners of neutrinos are heavy to contribute to the
potential.
  
For convenience, we define $X = R_\perp m_{3/2}$. We
begin by assuming neutrinos are Dirac. On the one hand, if $X=1$, then the gravitino tower 
(including the 0-mode) exactly cancels the bosonic towers except for
the first layer. Therefore, we are left with 9 bosonic degrees of
freedom against 12 fermionic ones. This implies that if $X=1$, the fermions will
always ``win'' at the end. The region of the parameter space which can
develop an AdS vacuum is determined by the mass of the neutrinos (and
of the radion). On the other hand, if $X=2$, then the gravitino tower 
(including the 0-mode) exactly cancels the bosonic towers except for the
first two layers. We therefore have 17 bosonic degrees of freedom against 12
fermionic ones. This implies that if $X=2$  the bosons will always
``win'' at the end and so the potential is unbounded from
below. Of course $X<1$, would also work and even relax the constraint on
$m_{1,{\rm max}}$ like in the analysis of Sec.~\ref{sec:5a}. Consequently, it seems that the interesting range of the
gravitino mass to compensate for the bosonic towers is $0\leqslant
X<2$.

In the case of three Majorana neutrinos however, we already know that
we need to add new light fermionic degrees of freedom to avoid the AdS
vacua. Therefore, for Majorana neutrinos the interesting range is
$0\leqslant X<1$. Using the results of
Tables~\ref{tab:radiongravitino} and \ref{tab:radiongravitinomaj} we
can obtain the maximal values of $X$ needed to avoid a stable AdS
vacuum. These values are given in Table~\ref{tab:X}. By comparing Tables
\ref{tab:radiongravitinomaj}  and \ref{tab:X} we conclude that in the
presence of a very light gravitino Majorana neutrinos can support stable dS vacua.

\begin{table}[!h]
\caption{Maximum $m_{3/2}$ and $X$ necessary to avoid an AdS
      vacuum.\protect\footnote{We have assumed that the radion and lightest
neutrino are massless.}}
    \begin{tabular}{p{0.3cm}p{0.3cm}||c|c|c||c|c|c}
      \hline
      \hline
    \multicolumn{2}{c||}{}&\multicolumn{3}{c||}{NO} & \multicolumn{3}{c}{IO}\\
    \cline{3-8}
    \multicolumn{2}{c||}{} &   $R_{\perp}$ & $m_{3/2}$ & $X$ &  $R_{\perp}$ & $m_{3/2}$ & $X$ \\
   \hline
        \hline
                           &
                             \parbox[t]{2mm}{\multirow{6}{*}{\rotatebox[origin=c]{90}{Dirac}}} &  ~~~~~5 $\mu$m~~~~~ & ~~~~~54.26 meV~~~~~ & ~~~~~1.375~~~~~ & ~~~~~5 $\mu$m~~~~~ & ~~~~~18.14 meV~~~~~ & ~~~~~0.460~~~~~\\
      &&  10 $\mu$m &22.35 meV& 1.133 & 10 $\mu$m &  10.15 meV& 0.514 \\
    & &   15 $\mu$m & 12.72 meV & 0.967 & 15 $\mu$m & 6.13 meV &0.466\\
     &&   20 $\mu$m& 8.71 meV & 0.883& 20 $\mu$m & 4.36 meV & 0.442\\
       && 25 $\mu$m &6.56 meV& 0.831 &  25 $\mu$m & 3.38 meV  & 0.428\\
     & &  30 $\mu$m& 5.25 meV & 0.798&30 $\mu$m & 2.77 meV & 0.421 \\
        \cline{2-8}
       &
         \parbox[t]{2mm}{\multirow{6}{*}{\rotatebox[origin=c]{90}{Majorana}}} &  5 $\mu$m &10.79 meV & 0.273 & 5 $\mu$m & 6.83 meV & 0.173\\
     & &  10 $\mu$m &8.08 meV& 0.409 & 10 $\mu$m & 3.63 meV & 0.184 \\
    & &   15 $\mu$m & 4.89 meV& 0.372& 15 $\mu$m & 2.17 meV  & 0.165 \\
   &  &   20 $\mu$m& 3.46 meV& 0.351 & 20 $\mu$m & 1.54 meV & 0.156\\
     &  & 25 $\mu$m & 2.67 meV& 0.338 &  25 $\mu$m & 1.19 meV & 0.151 \\
    &  &  30 $\mu$m& 2.17 meV & 0.330&30 $\mu$m & 0.98 meV  & 0.149 \\
     \hline
        \hline    
    \end{tabular}    
    \label{tab:X}
\end{table} 

\begin{figure}[!thb]
\postscript{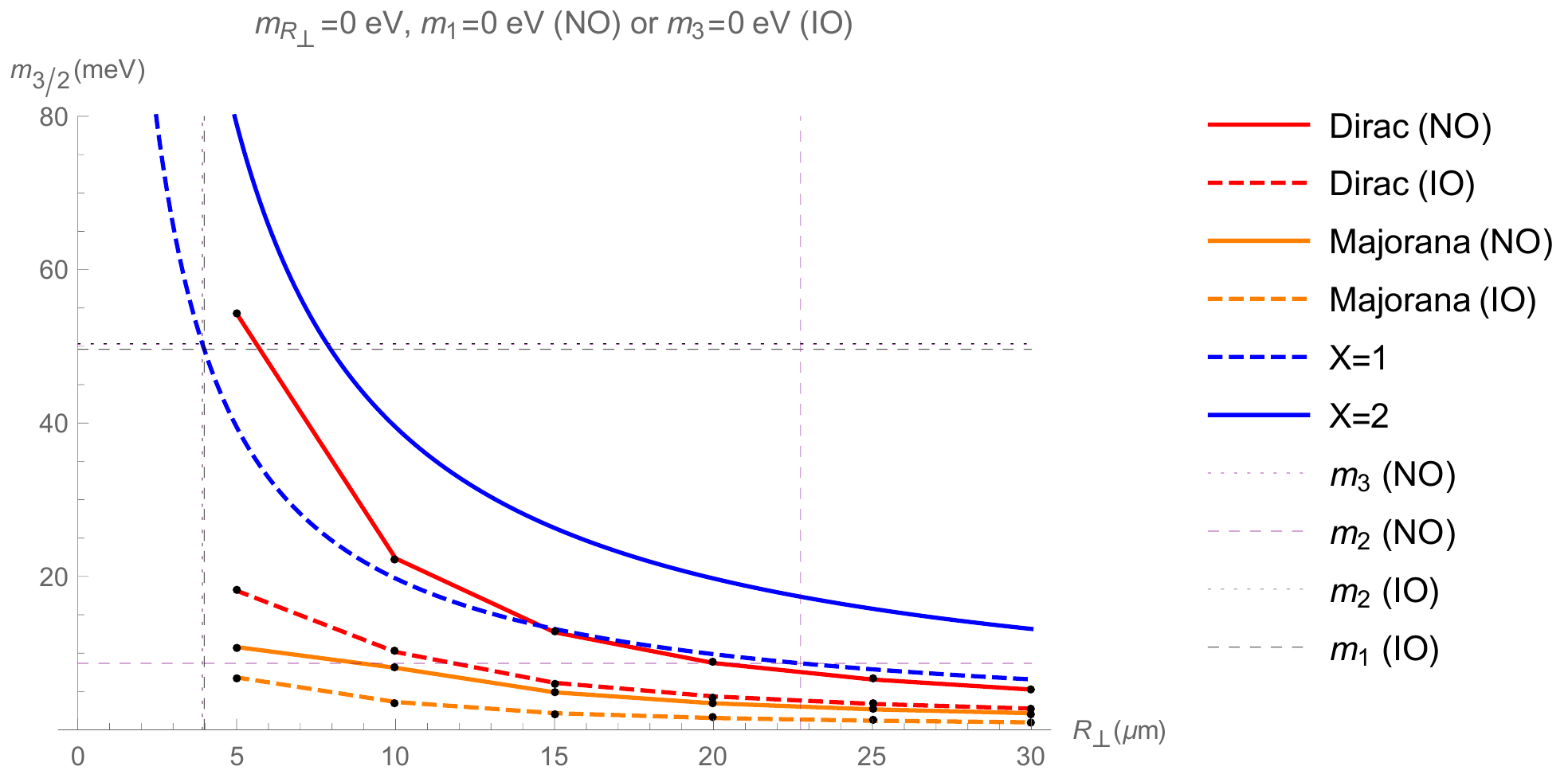}{0.8}
  \caption{Maximum gravitino mass needed to avoid the AdS vacuum.}
  \label{fig:onbrane}
 \end{figure}

In summary, if the gravitino is 
very light, then its KK tower can counterbalance the bosonic
towers to avoid AdS vacua in 4D $\to$ 3D compactifications. This
implies that the right-handed neutrinos and their left-handed
counterparts can be both localized on the brane. Besides, the 
gravitino 0-mode could help modifying the 3D Casimir vacua for the
case of Majorana neutrinos to become viable. The maximum gravitino
mass needed to avoid the AdS vacuum for Majorana and Dirac
neutrinos, assuming NO and IO, is summarized in Fig.~\ref{fig:onbrane}.

\section{Bulk neutrino masses}
\label{sec:6}

It has been pointed out that bulk neutrino masses allows relaxing the bounds on
$R_\perp$~\cite{Lukas:2000wn,Lukas:2000rg,Carena:2017qhd}. Considering
that fact, we now add Dirac masses for the three 5D neutrino fields.  For
simplicity, we swap to an intermediate mass basis $\Psi_i$ in which the 
flavor mixing has been already diagonalized. The kinetic and mass
terms in the Lagrangian take the form 
\begin{equation}
  \mathscr{L}   \supset \sum_{i=1}^3
\left[i\bar{\Psi}_i\Gamma^A
  \overset{\leftrightarrow}{\partial_{A}}\Psi_i-
  c_i\bar{\Psi}_i\Psi_i\right] \,,
\end{equation}
where $c_i$'s are the bulk mass parameters and $\Gamma^A = (\gamma^\mu, i \gamma^5)$. For convenience, we consider real $c_{i}$ and define
the mass matrix $\mathbb M_{i}$ by
\begin{align}
    \mathbb M_{i}=\begin{pmatrix}
        vY_{0}^{i} & 0 & \cdots & 0\\
        vY_{1}^{i} & m_{1}^{i} & \cdots & 0\\
        \vdots & 0 & \ddots & 0\\
        vY_{N}^{i} & 0 & \cdots & m_{N}^{i}
    \end{pmatrix} \,,
\end{align}
where 
\begin{equation}
    Y_{0}^{i}=y_{i}\sqrt{\frac{2}{\Lambda_{\rm QG}}}\sqrt{\frac{c_{i}}{e^{2c_{i}R_\perp\pi}-1}}
  \end{equation}
  and
  \begin{equation}
    Y_{n}^{i}=y_{i}\sqrt{\frac{2}{\Lambda_{\rm QG}\pi R_\perp}}\sqrt{\frac{n^{2}}{n^{2}+c_{i}^{2}R_\perp^{2}}}
\end{equation}
are the 4D Yukawa couplings (localized on the SM brane), with
$(m_{n}^{i})^{2}=(n/R_\perp)^{2}+c_{i}^{2}$~\cite{Carena:2017qhd}. We
can then compute
\begin{eqnarray} \lim\limits_{N\rightarrow\infty}\det\left(\mathbb
  M_{i}^{\dag} \mathbb
  M_{i}-\frac{\alpha_{i,n}^2}{R_\perp^{2}}\mathds{1}\right)& = &\left[\frac{c_{i}\xi_{i}^{2}}{e^{2c_{i}R_\perp\pi}-1}-\frac{\alpha_{i,n}^{2}}{R_\perp^{2}}+\frac{\xi_{i}^{2}}{2\pi
                                                               R_\perp} \right.
                                                               \nonumber \\
&\times & \left.
                                                               \left(\pi\sqrt{\alpha_{i,n}^{2}-c_{i}^{2}R_\perp^{2}}\cot\big(\pi\sqrt{\alpha_{i,n}^{2}-c_{i}^{2}R_\perp^{2}}\big)-c_{i}R_\perp\pi\coth\big(c_{i}R_\perp\pi\big)\right)\right]
  \nonumber \\
&\times & \prod\limits_{j=1}^{\infty}\left(\big(m_{j}^{i}\big)^{2}-\frac{\alpha_{i,n}^{2}}{R_\perp^{2}}\right)
  \nonumber \\
&=&\left[-\frac{c_{i}\xi_{i}^{2}}{2}-\frac{\alpha_{i,n}^{2}}{R_\perp^{2}}+\frac{\xi_{i}^{2}}{2R_\perp}\sqrt{\alpha_{i,n}^{2}-c_{i}^{2}R^{2}}\cot\big(\pi\sqrt{\alpha_{i,n}^{2}-c_{i}^{2}R_\perp^{2}}\big)\right]
    \nonumber \\
  & \times &
    \prod\limits_{j=1}^{\infty}\left(\big(m_{j}^{i}\big)^{2}-\frac{\alpha_{i,n}^{2}}{R_\perp^{2}}\right) \,,
\end{eqnarray}
where $\xi_{i}= v \ y_{i} \ \sqrt{2/\Lambda_{\rm QG}}$. Note
that the $\alpha_{i,n}^{2}/R_\perp^{2}$ solutions of this equation are also
the eigenvalues of $\mathbb M_{i} \mathbb M_{i}^{\dagger}$. In the
limit $c_{i}\rightarrow 0$, we recover from the bracket the usual transcendental equation:
\begin{align}
    \alpha_{i,n}-\pi\big(m_{i}^{\text{D}}R_\perp\big)^{2}\text{cot}\left(\pi\alpha_{i,n}\right)=0
  \, .
\end{align}
Note that the equation 
\begin{align}\label{LED+KK}
    -\frac{c_{i}\xi_{i}^{2}}{2}-\frac{\alpha_{i,n}^{2}}{R_\perp^{2}}+\frac{\xi_{i}^{2}}{2R_\perp}\sqrt{\alpha_{i,n}^{2}-c_{i}^{2}R_\perp^{2}}\cot\big(\pi\sqrt{\alpha_{i,n}^{2}-c_{i}^{2}R_\perp^{2}}\big)=0
\end{align}
has two different behaviors. If $(\alpha_{i,n}/R_\perp)^{2}\geqslant c_{i}^{2}$ it has an infinite number of solutions (the KK-tower) but if we look for solutions lighter than the mass in the bulk, namely $(\alpha_{i,n}/R_\perp)^{2}<c_{i}^{2}$, it becomes
\begin{align}\label{LED+0}
    -\frac{c_{i}\xi_{i}^{2}}{2}-\frac{\alpha_{i,n}^{2}}{R_\perp^{2}}+\frac{\xi_{i}^{2}}{2R_\perp}\sqrt{c_{i}^{2}R_\perp^{2}-\alpha_{i,n}^{2}}\coth\big(\pi\sqrt{c_{i}^{2}R_\perp^{2}-\alpha_{i,n}^{2}}\big)=0 \,,
\end{align}
which can have at most one solution. Actually, the function defined by
\begin{align}
    f(\alpha_{i,n})=-\frac{c_{i}\xi_{i}^{2}}{2}+\frac{\xi_{i}^{2}}{2R_\perp}\sqrt{c_{i}^{2}R_\perp^{2}-\alpha_{i,n}^{2}}\coth\big(\pi\sqrt{c_{i}^{2}R_\perp^{2}-\alpha_{i,n}^{2}}\big)
\end{align}
is monotonically decreasing on $\big[0,|c_i|R_\perp\big)$ whereas $\alpha_{i,n}^{2}/R_\perp^{2}$ is monotonically increasing. Moreover we have
\begin{align}
    f(0)=\frac{\xi_{i}^{2}}{2}\Big(|c_{i}|\coth(\pi|c_i|R_\perp)-c_{i}\Big)>\frac{\xi_{i}^{2}}{2}\Big(|c_{i}|-c_{i}\Big)\geqslant 0
    \end{align}
    and\begin{align}\lim\limits_{\alpha_{i,n}\rightarrow|c_{i}|R}f(\alpha_{i,n})=(m_{i}^{\text{D}})^{2}(1-\pi
         c_{i}R_\perp) \,,
\end{align}
so that equation \eqref{LED+0} has a unique solution if and only if 
\begin{align}
   (m_{i}^{\text{D}})^{2}(1-\pi c_{i}R_\perp) <  c_{i}^{2}\,;
\end{align}
namely, if 
\begin{align}
    c_{i}\notin\left[\frac{1}{2}(m_{i}^{\text{D}})^{2}\pi
  R_\perp\left(-1-\sqrt{1+\frac{4}{(m_{i}^{\text{D}})^{2}\pi^{2}R_\perp^{2}}}\right),\frac{1}{2}(m_{i}^{\text{D}})^{2}\pi
  R_\perp\left(-1+\sqrt{1+\frac{4}{(m_{i}^{\text{D}})^{2}\pi^{2}R_\perp^{2}}}\right)\right] \nonumber
\end{align}
and has no solution otherwise. This means that even if the spectrum is
shifted by adding a mass in the bulk, we can still have a light
0-mode. The eigenvectors of $\mathbb M_{i}^{\dagger}
\mathbb M_{i}$ are given by
\begin{align}
    \tilde{\mathbb
  V}_{i,\beta_n}=\tilde{\gamma}_{i,\beta_n}\begin{pmatrix}1&vY_{1}^{i}\frac{m_{1}^{i}}{\beta_n-(m_{1}^{i})^{2}}&vY_{2}^{i}\frac{m_{2}^{i}}{\beta_n-(m_{2}^{i})^{2}}&\cdots\end{pmatrix}^{\intercal} \,,
\end{align}
and the eigenvectors of $\mathbb M_{i} \mathbb M_{i}^{\dagger}$ are given by
\begin{align}
    \mathbb V_{i,\beta_n}=\mathbb M_{i}\tilde{\mathbb
  V}_{i,\beta_n}=\gamma_{i,\beta_n}\begin{pmatrix}vY_{0}^{i}&vY_{1}^{i}\left(1+\frac{(m_{1}^{i})^{2}}{\beta_n-(m_{1}^{i})^{2}}\right)&vY_{2}^{i}\left(1+\frac{(m_{2}^{i})^{2}}{\beta_n-(m_{2}^{i})^{2}}\right)&\cdots\end{pmatrix}^{\intercal}\, ,
\end{align}
where $\beta_n=\alpha_{i,n}^{2}/R_\perp^{2}$ is the associated eigenvalue
and where $\tilde{\gamma}_{i,\beta_n}$ and $\gamma_{i,\beta_n}$ are
normalization factors. We are interested in eigenvectors satisfying
$\tilde{\mathbb V}_{i,\beta_n}^{\intercal}\tilde{\mathbb V}_{i,\beta_n}= \mathds{1}$ 
 and $\mathbb V_{i,\beta_n}^{\intercal} \mathbb V_{i,\beta_n}=
 \mathds{1}$, which lead to
\begin{equation}
\tilde{\gamma}_{i,\beta_n}^{2} =\frac{2\Lambda_{\rm QG}}{2\Lambda_{\rm
                              QG}+v^{2}
                            y_i^{2}\left[-\frac{\cot\Big(\pi
                                R_\perp\sqrt{\beta_n-c_{i}^{2}}\Big)}{\sqrt{\beta_n-c_{i}^{2}}}+\pi
                              R_\perp\csc^{2}\Big(\pi
                              R_\perp\sqrt{\beta_n-c_{i}^{2}}\Big)\right]},
                        \end{equation}
                        and
                        \begin{eqnarray}
\gamma_{i,\beta_n}^{2} & = &\frac{4\Lambda_{\rm
                      QG}\sqrt{\beta_n-c_{i}^{2}}}{v^{2}
                          y_{i}^{2}\Big[\csc\big(\pi
                          R_\perp\sqrt{\beta_n-c_{i}^{2}}\big)\Big]^{2}}  
                        \left[2\sqrt{\beta_n-c_{i}^{2}}(\pi
                           R_\perp\beta_n-c_{i})+2c_{i}\sqrt{\beta_n-c_{i}^{2}}\right. \nonumber \\
                      &\times & \left. \cos\big(\pi
                           R_\perp\sqrt{\beta_n-c_{i}^{2}}\big) 
+ (\beta_n-2c_{i}^{2})\sin\big(\pi R_\perp\sqrt{\beta_n-c_{i}^{2}}\big)\right]^{-1}.
\end{eqnarray}
With this in mind, the masses and mixing between the 0-mode of neutrino species $i$ and
the $n^{\rm th}$ KK-mode are characterized by\footnote{Note that contrary to what is stated in~\cite{Carena:2017qhd}, the left rotation that
diagonalizes the mass matrix $\mathbb{M}_i$ in the intermediate basis
relates to $\mathbb M_{i}^{\dagger} \mathbb M_{i}$
and rather than $\mathbb M_{i} \mathbb M_{i}^{\dagger}$.} 
\begin{align}
    \left|L^{0n} \right|^{2}=\tilde{\gamma}_{i,\beta_{n}}^{2},\qquad
  n\in\mathds{N} \, .
\end{align}
Using \eqref{LED+KK} we can express $\tilde{\gamma}_{i,\beta_n}^{2}$ as 
\begin{align}
    \tilde{\gamma}_{i,\beta_n}^{2}=\frac{8\big(m_{i}^{D} \big)^2\pi
  R_\perp(c_{i}^{2}-\beta_n)}{4c_{i}\big(m_{i}^{D}\big)^{4}\pi^{2}R_\perp^{2}+8\big(m_{i}^{D}\big)^{2}\pi
  R_\perp(c_{i}^{2}-\beta_n)- {\cal A} -   4\pi
  R_\perp\beta_n^{2}} \, .
\end{align}
with ${\cal A} = 2\big(m_{i}^{D}\big)^{2}\pi R_\perp\Big\{-2+\pi
  R_\perp \big[4c_{i}+2\big(m_{i}^{D}\big)^{2}\pi
  R_\perp\big]\Big\}\beta_n$.

\begin{figure}[htb!]
    \postscript{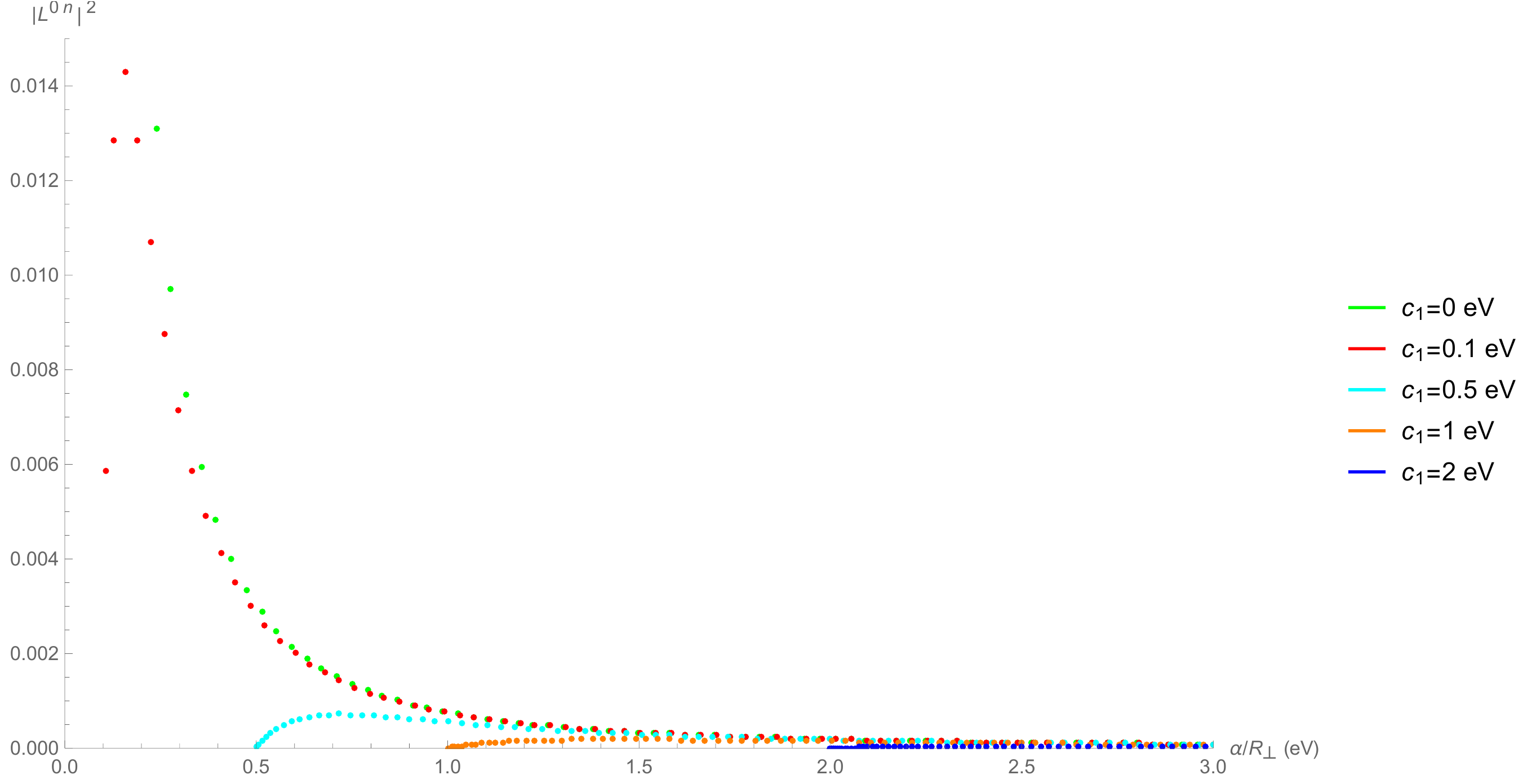}{0.92}
  \caption{Pattern of KK masses and mixings for  the
    lightest 0-mode assuming $y_1 = 0.001$ and fiducial values of
    $c_i$ and $R_\perp = 5\mu{\rm m}$.} \label{fig:LED+}
\end{figure}

\begin{figure}[htb!]
    \postscript{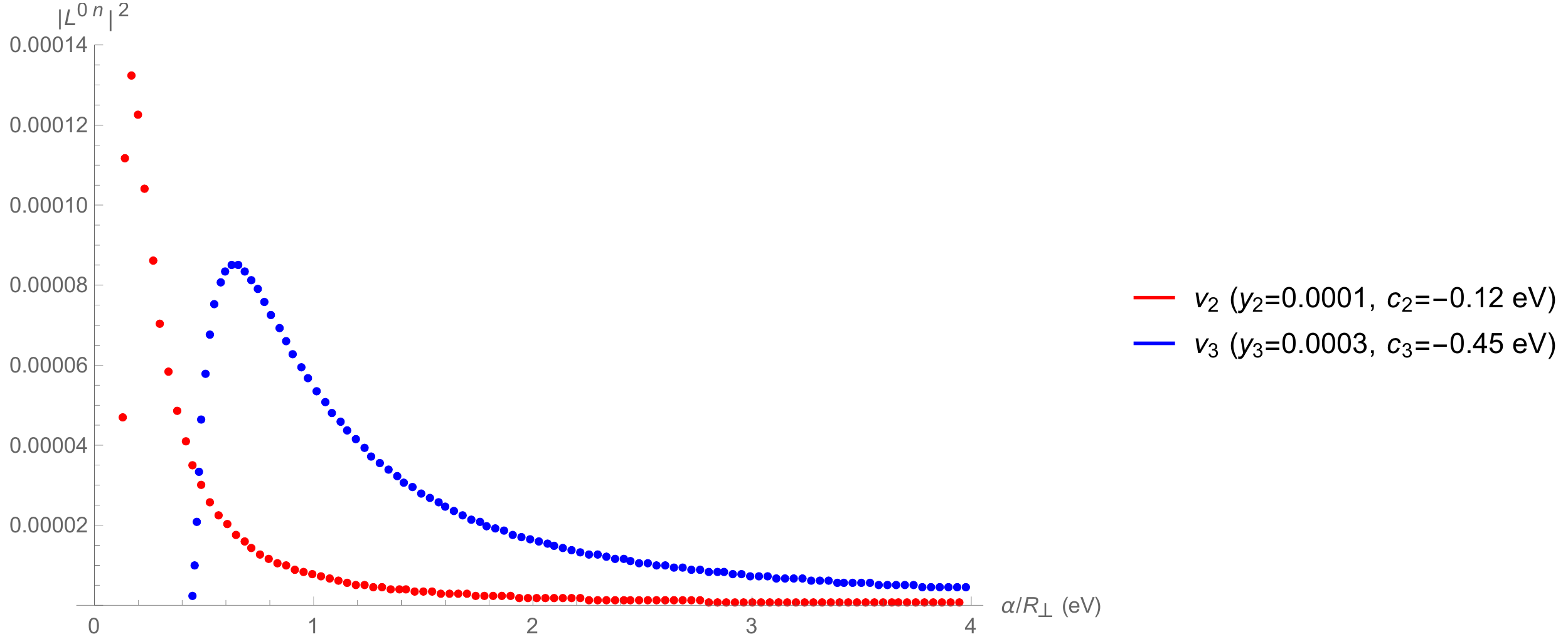}{0.92}
    \caption{Pattern of KK masses and mixings for the heaviest zero
      modes assuming NO and a compactification radius $R_\perp = 5\mu{\rm m}$. We have taken $c_2
    = -0.12~{\rm eV}$,  $c_3 = -0.45~{\rm eV}$, $y_2 = 0.0001$, and  $y_3 = 0.0003$.} \label{fig:LED+2}
\end{figure}

In Fig.~\ref{fig:LED+} we show possible examples of the oscillation
pattern of KK masses and mixings for fiducial values $c_i$,
$y_i$, and $R_\perp$. We can see that for $c_1 = 0.1~{\rm eV}$ there
is a strong suppression of the mixing with the first 2 KK modes, and $|L^{0n}|$ peaks
is at $\alpha_{1,3}/R_\perp = 156~{\rm meV}$, with
$|L^{03}|^2 = 0.014$. This is in sharp contrast with the case for $c_1
=0$, in which  $|L^{0n}|$ peaks in the first KK at $\alpha_{1,1}/R_\perp =
46~{\rm meV}$, with $|L^{01}|^2 = 0.22$. For higher values of $c_1$,
there is
suppression of a larger number of KK modes and at the peak
$|L^{0n}|^2$ becomes even smaller. For the neutrino towers to be able
to compensate the bosonic towers, the mass of the lightest neutrino
0-mode should be very small, e.g., for $c_1 = 0.1~{\rm eV}$ and $R_\perp =
5~\mu{\rm m}$  we have $\alpha_{1,0}/R_\perp \sim 2.5 \times 10^{-2}~{\rm
  meV}$. As shown in Fig.~\ref{fig:LED+2}, the second
and third 0-modes also have a strong suppression on the mixing with
the first  2 and 10 KK modes, respectively. For $\nu_2$ and $\nu_3$,
$|L^{0n}|$ peaks in the third and eleventh KK mode at $\alpha_{2,3}/R_\perp =
169~{\rm meV}$  and $\alpha_{3,11}/R_\perp = 625~{\rm meV}$, with $|L^{03}|^2 = 0.00013$
and $|L^{0\,11}|^2 = 0.0000849$. This corresponds to $\Delta m^2_{21}
\sim 7.2 \times 10^{-5}~{\rm eV}^2$, $\Delta m^2_{32} \sim 2.4 \times
10^{-3}~{\rm eV}^2$, and $\sum m_\nu \sim 58~{\rm meV}$ in good agreement
with observations. See Appendix for further details.

Now, for $R_\perp \gtrsim 10~\mu {\rm m}$, the masses of the bosonic
KK become of the order of the neutrino 0-modes, and therefore close to
$|c|$ if we need to enforce a suppression of 0-mode oscillations into
the first few KK states. If this were the case, the neutrino towers
(whose masses can be approximated by $\sqrt{(n/R_\perp)^2+c_i^2}$, $n>0$) would be
significantly shifted from the bosonic towers (whose masses can be
approximated by $n/R_\perp$). This implies that the neutrino towers would not
be sufficient to cancel the influence of the bosonic towers
in carving $V(R)$. But again, to balance the bosonic towers a very light
gravitino may come to the rescue.

\section{Conclusions}
\label{sec:7}

The Swampland program has made the striking proposal  that if the low-energy effective theory is the minimal SM extension accommodating neutrino masses, then
  neutrinos cannot be Majorana particles. This is because the {\it
    sharpened} version of the weak gravity conjecture forbids the
  existence of non-SUSY AdS vacua supported by fluxes in a consistent
  quantum gravity  theory, and if neutrinos are Majorana, when
  the SM + GR are compactified down
  to 3D, then AdS vacua appear for any values of neutrino
  masses consistent with experiment. However, this is not the case if
  neutrinos are Dirac particles, for which the SM + GR
  compactification down to 3D sets a limit on the required maximum
  mass of the lightest neutrino to carry dS rather than AdS vacua. Motivated by these astonishing
  results we have studied the landscape of lower-dimensional vacua
  that arise in the SM coupled to gravity enriched with the dark dimension. The results of our
  investigation can be summarized as follows:
\begin{itemize}
\item If righ-handed neutrinos propagate in the bulk  (so
  that their Yukawa couplings become tiny due to a volume
  suppression) then their KK towers can compensate for the
  graviton tower to avoid AdS vacua. However, data from
  neutrino oscillation experiments set restrictive bounds on the
  compactification radius and so the first KK
  neutrino modes are too heavy to alter the shape of the radion
  potential or the maximum mass of the lightest neutrino state from
  those predicted by the SM + GR when compactified down to 3D.
\item A very light gravitino (with mass in the meV range) could help
  relaxing the neutrino mass constraint. The difference between the predicted
  total neutrino mass $\sum m_\nu$ by SM + GR and SM + GR in the
  presence of a very light gravitino propagating through the dark dimension is within reach of next-generation cosmological probes that will
  measure the total neutrino mass with an uncertainty
  $\sigma (\sum m_\nu) = 0.014~{\rm eV}$.
\item If the gravitino is very light, then its 
  KK tower can compensate for the graviton
  tower to avoid AdS vacua and thus right-handed neutrinos can
  (in principle) be locked on the brane. For this scenario,
  Majorana neutrinos could develop dS vacua.
\item Bulk neutrino masses 
  can suppress the mixing with the first KK mode in the neutrino towers and relax the
  oscillation bound on the compactification radius,  but at the expense of shifting the
  KK neutrino towers to higher masses. However, there is a neutrino
  0-mode that
  can stay light in each tower to accommodate neutrino oscillation
  data and the cosmological bound.
\end{itemize}
As a by product of our investigation focusing on the validity of the non-SUSY AdS instability 
conjecture we end up with a prediction of Swampland phenomenology within the framework of the dark
dimension: {\it either the gravitino is very light or else neutrinos have to
  be Dirac with the right-handed states propagating into the bulk, so
  that the neutrino towers can compensate the contribution of the
  graviton KK modes to the potential.}

\section*{Acknowledgments}

We have greatly benefited from discussions with Nima
Arkani-Hamed. L.A.A. is supported by the U.S. National Science
Foundation (NSF) Grant PHY-2112527.

\section*{Appendix}

\begin{figure}[b]
    \postscript{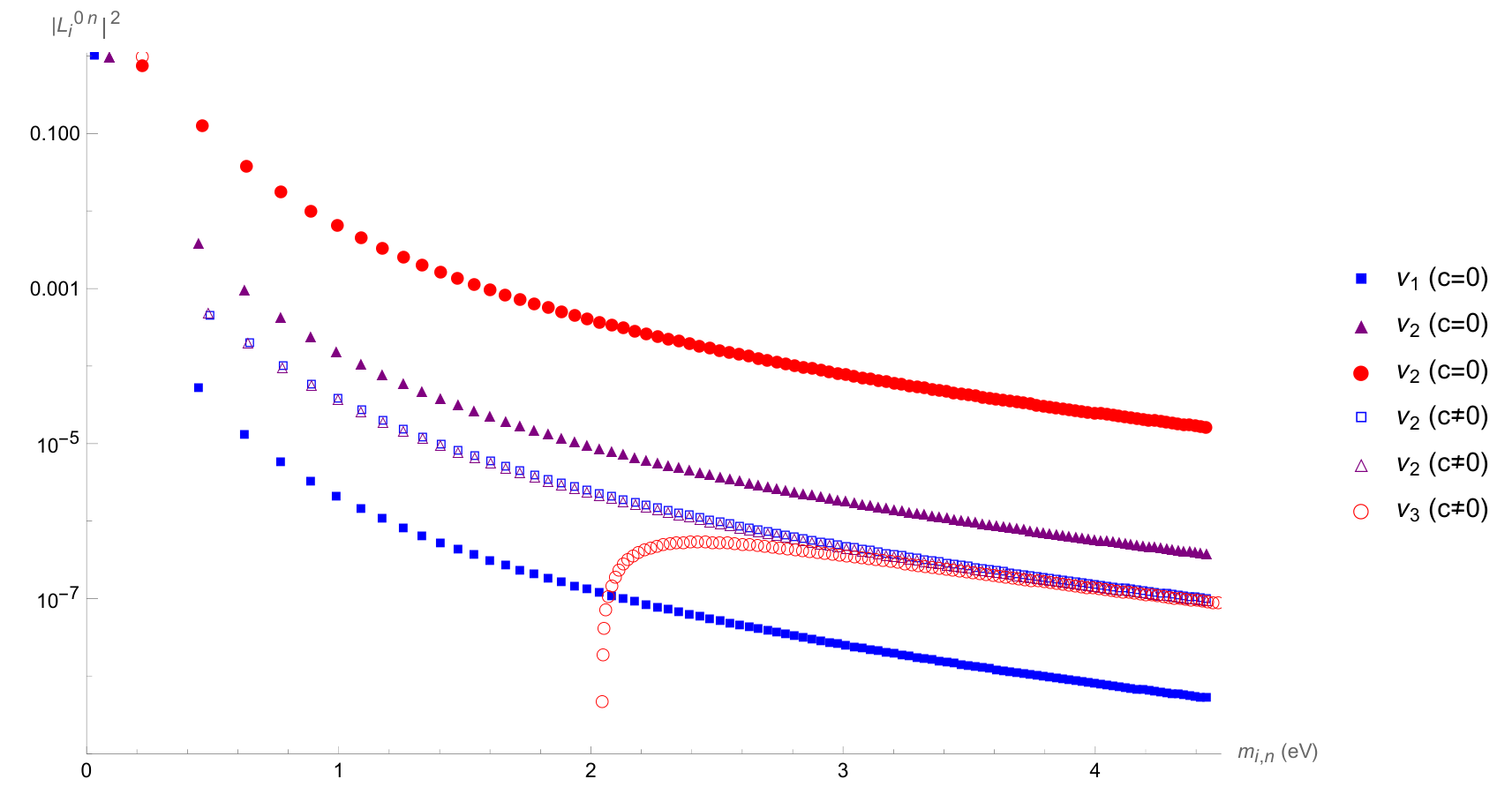}{0.8}
    \caption{Neutrino spectrum.} \label{fig:app1}
\end{figure}

\begin{figure}[htb!]
    \postscript{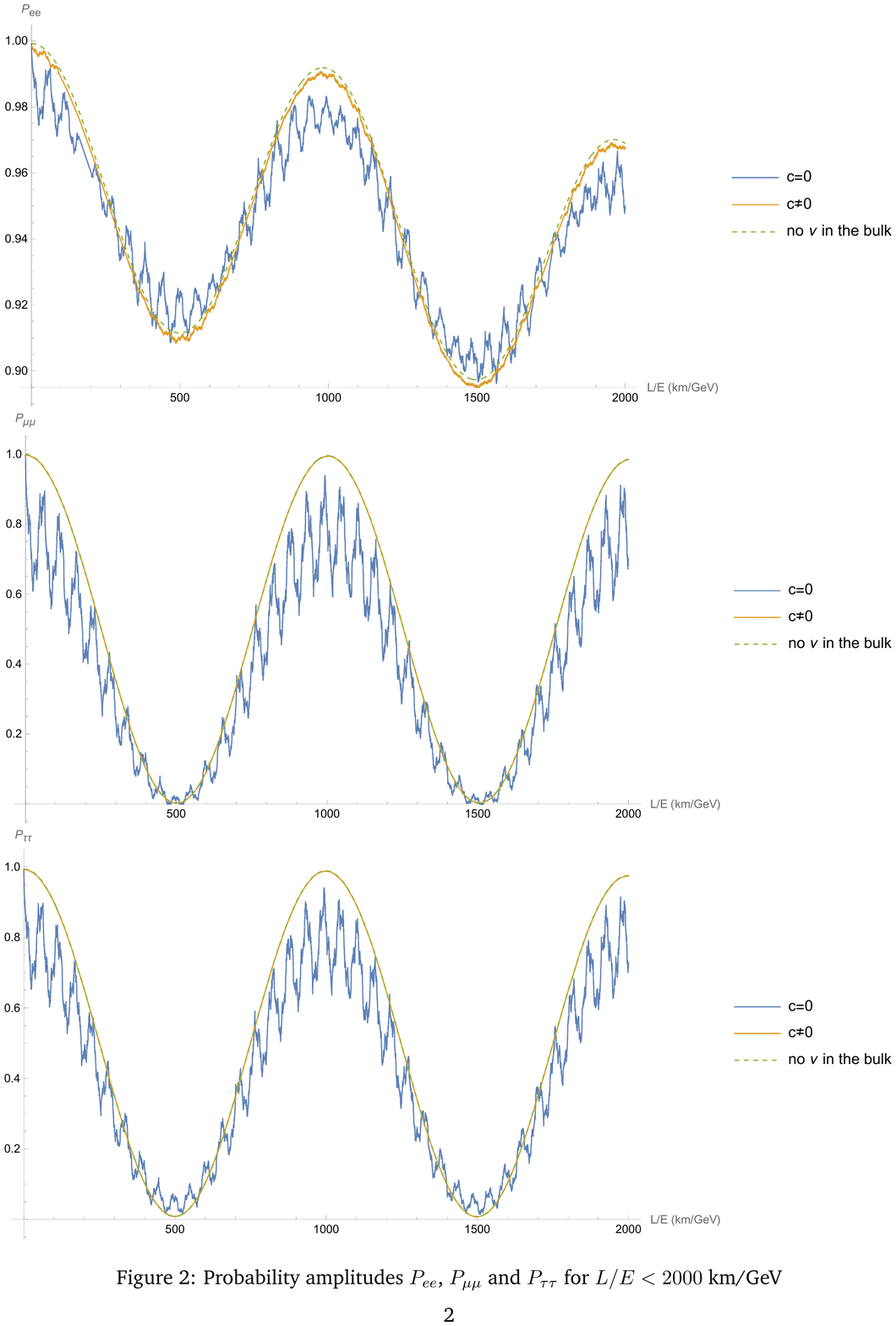}{0.8}
    \caption{Probability amplitudes $P_{ee}$, $P_{\mu \mu}$, and
      $P_{\tau \tau}$ for $L/E < 2000~{\rm km/GeV}$.  \label{fig:app2}}
\end{figure}

\begin{figure}[htb!]
    \postscript{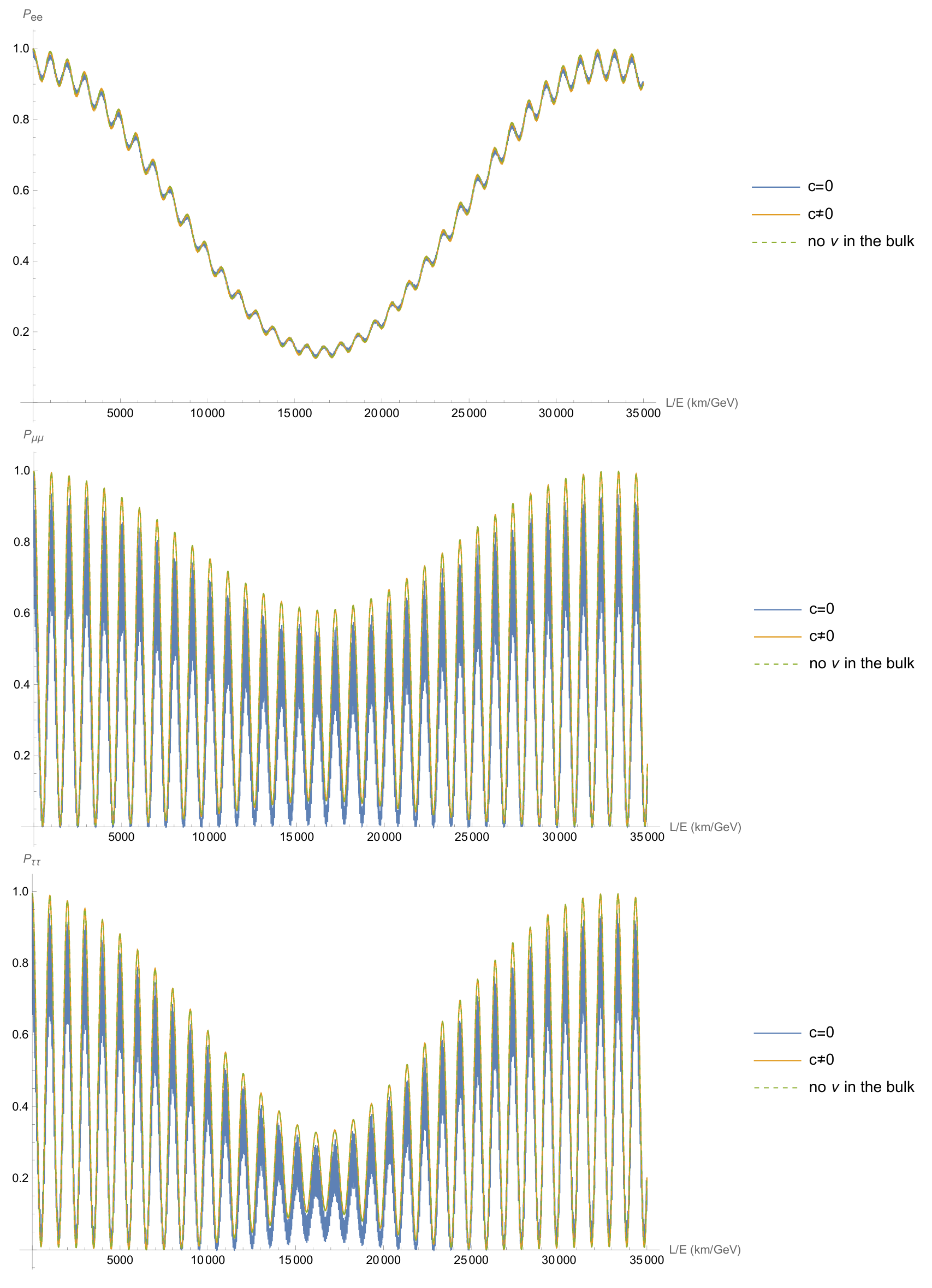}{0.8}
    \caption{Probability amplitudes $P_{ee}$, $P_{\mu \mu}$, and
      $P_{\tau \tau}$ for $L/E < 35000~{\rm km/GeV}$.  \label{fig:app3}} 
\end{figure}

\begin{figure}[htb!]
    \postscript{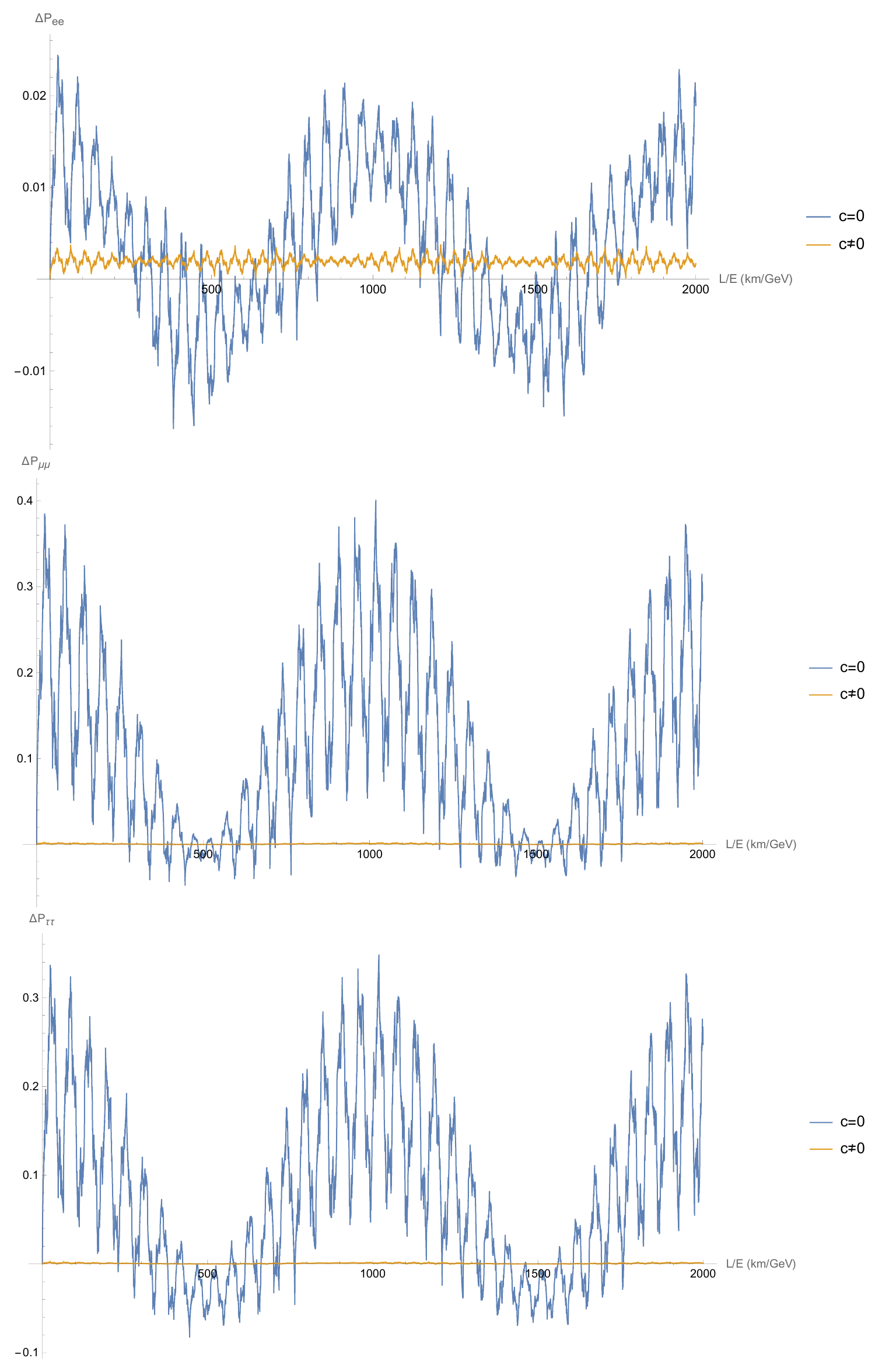}{0.8}
    \caption{Deviations of the probability amplitudes from the
      scenario without bulk neutrinos for $L/E <
      2000~{\rm km/GeV}$ \label{fig:app4}.} 
\end{figure}
\begin{figure}[htb!]
    \postscript{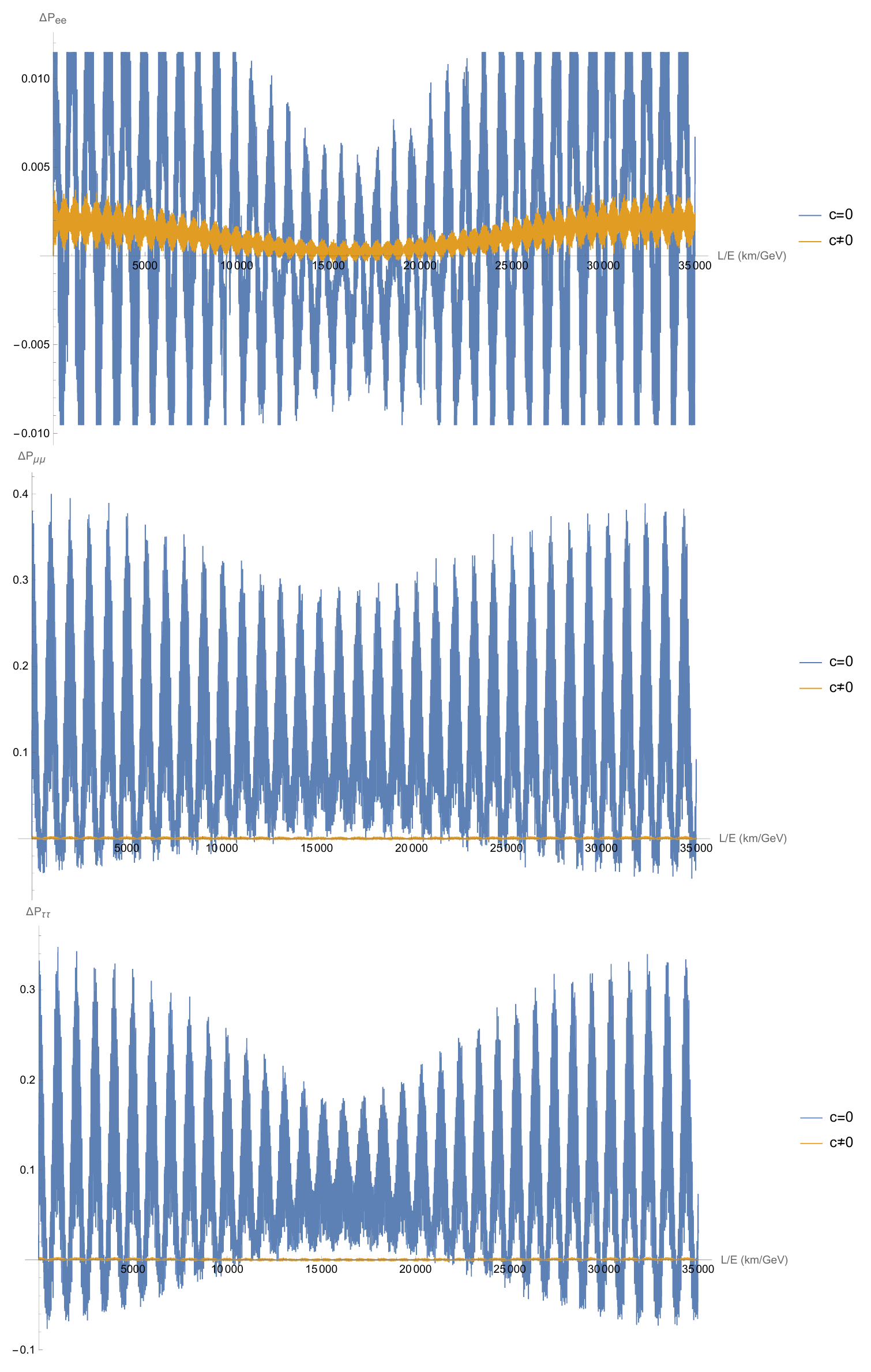}{0.79}
    \caption{Deviations of the probability amplitudes from the
      scenario without bulk neutrinos for $L/E <
      35000~{\rm km/GeV}$ \label{fig:app5}.} 
\end{figure}

For completness, in this Appendix we calculate the survival
probability in the presence of bulk masses. In order to be able to
compare the cases with and without bulk masses and clearly see the
strong suppression of the mixing with the first few KK modes, we have to look at a value of $R_\perp$ in which the scenario without
bulk masses is excluded. Indeed, because of the constraint from
oscillation data, it is not
possible to build a coherent spectrum without bulk masses for $R_\perp
\alt 0.4~\mu {\rm m}$ for NO~\cite{Anchordoqui:2022svl}. We will therefore build an explicit
example for NO with $R_\perp = 1~\mu{\rm m}$, and a choice of realistic and simple parameters. Namely, we choose $m_1^{(0)}
= 1~{\rm meV}$ which together with the value of $R_\perp$ fully determine the Yukawa
couplings, and then the whole spectrum, when there is no masses in
the bulk. Namely, it gives $y_1 \simeq 2 \times 10^{-5}$, $y_2 \simeq 2 \times 10^{-4}$ and $y_3 \simeq
1.3 \times 10^{-3}$, or equivalently $m^D_1 \simeq m_1^{(0)} \simeq
1~{\rm meV}$, $m^D_2 \simeq 9~{\rm meV}$ and $m^D_3 \simeq 57~{\rm meV}$. When we turn on the masses in the bulk, the choice of $R_\perp$ and of
the values of the zero mode masses do not determine the spectrum as
we still have two correlated parameters $c_i$ and $y_i$ for each $i$. Here we
 make the simple choice of $y_1 = y_2 = y_3 = 10^{-4}$. The bulk masses
are thus determined and are found to be $c_1 \simeq 140~{\rm meV}$,
$c_2 \simeq -124~{\rm meV}$
and $c_3 \simeq -4.177~{\rm  eV}$. The spectrum is therefore fully determined and is
presented in Fig.~\ref{fig:app1}.

We can now compare the survival probability in the three cases: no
neutrino in the bulk (SM case), and neutrinos in the bulk with and
without bulk masses, all as functions of $L/E$, where $L$ is the
experiment baseline, $E$ is the neutrino energy. To this end, we first
define the relation between the flavor and intermediate bases
\begin{equation}
  \nu_{\alpha,0}^L = U_{\alpha i}\,\nu_{i,0}^L,\quad 
  \Psi_{\alpha}=R_{\alpha i}\Psi_{i} \,,
  \label{eq:rotate1}
\end{equation}
where $U_{\alpha i}$ is the usual Pontecorvo--Maki--Nagakawa--Sakata 
matrix~\cite{Pontecorvo:1967fh,Pontecorvo:1957qd,Maki:1962mu} for the
standard three flavor neutrino model and $R$ is a matrix that
diagonalizes the bulk masses and Yukawa couplings. The oscillation
amplitude (in vacuum) among active neutrinos is given by
\begin{equation}\label{eq:amplitude}
  \mathcal{A}(\nu_{\alpha,0}\to\nu_{\beta,0}; L) = \sum_{i,n}\mathcal{U}_{\alpha i}^{0n}(\mathcal{U}_{\beta i}^{0n})^*\exp\left(i \frac{ m_{i,n}^2 L}{2E}\right),
\end{equation}
where $\mathcal{U}_{\alpha i}^{0n} = U_{\alpha i}L_i^{0n}$, and where
and the superscripts indicating left-handedness have been dropped. It
is noteworthy that the other entries of $\mathcal{U}_{\alpha i}^{nm}$~\cite{Carena:2017qhd}
are not observable, because the sterile neutrinos do not couple to the
electroweak gauge bosons. The survival probability of flavor
$\alpha$ a distance $L$ is given by
\begin{equation}
P_{\alpha \alpha} \equiv P(\nu_{\alpha,0} \to \nu_{\alpha,0}) =
  \left|\mathcal{A}(\nu_{\alpha,0}\to\nu_{\alpha,0}; L) \right|^2 \, .
\end{equation}  
In Figs.~\ref{fig:app2} and \ref{fig:app3} we show the survival
probability of the different flavors as a function of $L/E$ showing different scales. We
can see that the case with bulk masses is way more similar to the SM
case than the case without bulk masses. In order to refine this
statement, in Figs.~\ref{fig:app4} and \ref{fig:app5} we show the
differences of probabilities with respect to the SM (without bulk
neutrinos) scenario. We can see that even though the case without bulk
masses has huge differences with respect to the SM case, this is no
longer true once we include bulk masses. Moreover, we remind the
reader that when we include bulk masses, choosing $R_\perp$ and the
masses of the zero modes do not fully determine the spectrum. We
decided to display simple and generic values of the parameters here to
see the natural behavior of the system, but it is definitely possible
to adjust the values of $y_i$ and $c_i$ to have the same zero modes
spectrum, but to reduce even more the effect of the KK towers.

\end{document}